\documentclass[reqno,12pt,a4paper]{article}
\pdfoutput=1
\usepackage{dpdefs}
\usepackage{tweaklist}
\usepackage{amsmath}
\usepackage{amssymb}
\usepackage{amsfonts}
\usepackage{graphicx}
\providecommand{\U}[1]{\protect\rule{.1in}{.1in}}
\renewcommand{\enumhook}{\setlength{\topsep}{1ex} \setlength{\itemsep}{-0.2ex}}
\let\oldmarginpar\marginpar
\renewcommand\marginpar[1]{\-\oldmarginpar[\raggedleft\footnotesize #1]{\raggedright\color{blue}\singlespacing\footnotesize #1}}
\newtheorem{theorem}{Theorem}
\newtheorem{lemma}[theorem]{Lemma}
\newtheorem{assumption}[theorem]{Assumption}
\newtheorem{proposition}[theorem]{Proposition}
\newtheorem{corollary}[theorem]{Corollary}
\hypersetup{pdfstartview={FitH}, citecolor=blue }
\floatstyle{ruled}
\newfloat{table}{htbp}{lol}
\floatname{table}{Table}
\begin{document}

\title{Higher-Order Improvements of the Sieve Bootstrap for Fractionally Integrated
Processes\thanks{This research has been supported by Australian Research
Council (ARC) Discovery Grants DP0985234 and DP120102344, and ARC Future
Fellowship FT0991045. The authors would like to thank the Editor and two
referees for very detailed and constructive comments on an earlier draft of
the paper.}}
\author{D. S. Poskitt, Simone D. Grose and Gael M. Martin\thanks{Corresponding author:
Gael Martin, Department of Econometrics and Business Statistics, Monash
University, Clayton, Victoria 3800, Australia. Tel.: +61-3-9905-1189; fax:
+61-3-9905-5474; email: gael.martin@monash.edu.}
\and \emph{{\small {Department of Econometrics \& Business Statistics, Monash
University}}}}
\maketitle

\begin{abstract}
This paper investigates the accuracy of bootstrap-based inference in the case
of long memory fractionally integrated processes. The re-sampling method is
based on the semi-parametric sieve approach, whereby the dynamics in the
process used to produce the bootstrap draws are captured by an autoregressive
approximation. Application of the sieve method to data pre-filtered by a
semi-parametric estimate of the long memory parameter is also explored.
Higher-order improvements yielded by both forms of re-sampling are
demonstrated using Edgeworth expansions for a broad class of statistics that
includes first- and second-order moments, the discrete Fourier transform and
regression coefficients. The methods are then applied to the problem of
estimating the sampling distributions of the sample mean and of selected
sample autocorrelation coefficients, in experimental settings. In the case of
the sample mean, the pre-filtered version of the bootstrap is shown to avoid
the distinct underestimation of the sampling variance of the mean which the
raw sieve method demonstrates in finite samples, higher order accuracy of the
latter notwithstanding. Pre-filtering also produces gains in terms of the
accuracy with which the sampling distributions of the sample autocorrelations
are reproduced, most notably in the part of the parameter space in which
asymptotic normality does not obtain. Most importantly, the sieve bootstrap is
shown to reproduce the (empirically infeasible) Edgeworth expansion of the
sampling distribution of the autocorrelation coefficients, in the part of the
parameter space in which the expansion is valid.

\end{abstract}



\titlepage

\setcounter{footnote}{0}

\section{Introduction}

Many empirical time series have been found to exhibit behaviour characteristic
of long memory, or long-range dependent, processes, and the class of
fractionally integrated ($I(d)$) processes introduced by
\cite{granger:joyeux:1980} and \cite{hosking:1981} is perhaps the most popular
model used to describe the features of such processes. $I(d)$ processes can be
characterized by the specification
\begin{equation}
y(t)=\sum_{j=0}^{\infty}k(j)\varepsilon(t-j)=\frac{\kappa(z)}{(1-z)^{d}%
}\,\varepsilon(t), \label{Wold}%
\end{equation}
where $\varepsilon(t)$, $t\in{\mathcal{Z}}$, is a zero mean white noise
process with variance $\sigma^{2}$, $z$ is here interpreted as the lag
operator $(z^{j}y(t)=y(t-j))$, and $\kappa(z)=\sum_{j\geq0}\kappa(j)z^{j}$.
The behaviour of this process naturally depends on the fractional integration
parameter $d$; for instance, if the \textquotedblleft
non-fractional\textquotedblright\ component $\kappa(z)$ is the transfer
function of a stable, invertible autoregressive moving-average (ARMA) and
$|d|<0.5$, then the coefficients of $k(z)$ are square-summable, $\sum_{j\geq
0}|k(j)|^{2}<\infty$, and $y(t)$ is well-defined as the limit in mean square
of a covariance-stationary process. More pertinently, for any $d>0$ the
impulse response coefficients of $k(z)$ in the representation \eqref{Wold} are
not absolutely summable and the autocovariances decline at a hyperbolic rate,
$\gamma(\tau)\sim C\tau^{2d-1}$, rather than the exponential rate typical of
an ARMA process. For a detailed description of the properties of long memory
processes see \cite{beran:1994}.

Statistical procedures for analyzing fractional processes are discussed in
\cite{hosking:1996}, and techniques for estimating fractional models have
ranged from the likelihood-based methods studied in \cite{fox:taqqu:1986},
\cite{dahlhaus:1989}, \cite{sowell:1992} and \cite{beran:1995}, to the
semi-parametric methods advanced by \cite{geweke:porter:1983} and
\citet{robinson:1995a,robinson:1995b}, among others. These techniques
typically focus on obtaining an accurate estimate of the parameter governing
the long-term behaviour of the process, and the asymptotic theory for these
estimators is well established. In particular, we have consistency, asymptotic
efficiency, and asymptotic normality for the maximum likelihood estimator
(MLE), and the semi--parametric estimators are consistent and asymptotically
pivotal with particularly simple asymptotic normal distributions.

Concurrent with the development of the asymptotic theory associated with the
estimation of long memory models, focus has also been directed at the
production of more accurate estimates of finite sample distributions in this
setting. An explicit form for the Edgeworth expansion for the sample
autocorrelation function of a stationary Gaussian long memory process is
derived in \cite{lieberman:rousseau:zucker:2001}, and
\cite{lieberman:rousseau:zucker:2003} establish the validity of an Edgeworth
expansion for the distribution of the MLE of the parameters of such a process,
with a zero mean assumed. The unknown mean case is covered in
\cite{andrews:lieberman:2005}, with the estimator defined by maximizing the
log-likelihood with the unknown mean replaced by the sample mean (referred to
as the \textquotedblleft plug-in\textquotedblright\ MLE, or PML).
\cite{andrews:lieberman:2005} also derive results for the Whittle MLE (WML)
and for the plug-in version (PWML). \cite{giraitis:robinson:2003} derive an
Edgeworth expansion for the semi-parametric local Whittle estimator of the
long memory parameter \citep{robinson:1995b} (SPLW), whilst
\cite{lieberman:phillips:2004} derive an explicit form for the first-order
expansion for the MLE of the long memory parameter in the fractional noise case.

From the point of view of practical implementation, evaluation of the terms in
such expansions, for general long memory models, is no trivial task and
typically requires knowledge of the values of population ensemble parameters.
These expansions are also usually only valid under more restrictive
assumptions than are required for first-order asymptotic approximations; see,
for example, \cite{lieberman:rousseau:zucker:2001} and
\cite{giraitis:robinson:2003}. Accordingly, much attention has also been given
to the application of bootstrap-based inference in these models. Building on
the Edgeworth results of \cite{lieberman:rousseau:zucker:2003} and
\cite{andrews:lieberman:2005}, \cite{andrews:lieberman:marmer:2006} derive the
error rate for the parametric bootstrap for the PML and PWML estimators in
Gaussian autoregressive fractionally integrated moving average (ARFIMA)
models. In contrast, \cite{poskitt:2008} proposes a semi-parametric approach,
based on the sieve bootstrap, and provides both theoretical and
simulation-based results regarding the accuracy with which the method
estimates the true sampling distribution of suitably continuous linear
statistics. To the authors' knowledge \cite{andrews:lieberman:marmer:2006} and
\cite{poskitt:2008} are amongst the earliest papers in the literature to have
examined the theoretical properties of bootstrap methods in the context of
fractionally integrated (long memory) processes.

The current paper builds upon the results presented in \cite{poskitt:2008} and
produces new results regarding error rates for sieve-based bootstrap
techniques in the context of fractionally integrated processes. Using
Edgeworth expansions, it is shown that the procedure we here refer to as the
\textquotedblleft raw\textquotedblright\ sieve bootstrap can achieve an error
rate of $O_{p}(T^{-(1-d^{\prime})+\beta})$ for all $\beta>0$ where $d^{\prime
}=\max\{0,d\}$, for a class of statistics that includes the sample mean, the
sample autocovariance and autocorrelation functions, the discrete Fourier
transform and ordinary least squares (OLS) regression coefficients. We also
present a new methodology based on a modified form of the sieve bootstrap. The
modification uses a consistent semi-parametric estimator of the long memory
parameter to pre-filter the raw data, prior to the application of a long
autoregressive approximation which acts as the \textquotedblleft
sieve\textquotedblright\ from which bootstrap samples are produced. We refer
to this as the pre-filtered sieve bootstrap. We establish that, subject to
appropriate regularity, for any fractionally integrated processes with
$|d|<0.5$ the error rate of the pre-filtered sieve bootstrap is $O_{p}%
(T^{-1+\beta})$ for all $\beta>0$. These results generalize those of
\cite{choi:hall:2000} who show that, for linear statistics characterized by
polynomial products, double sieve bootstrap calibrated percentile methods and
sieve bootstrap percentile $t$ confidence intervals evaluated in the short
memory case converge at a rate arbitrarily close to that obtained with simple
random samples, namely $O_{p}(T^{-1+\beta})$ for all $\beta>0$.

\cite{choi:hall:2000} argue that for short memory processes the sieve
bootstrap is to be preferred over the block bootstrap \citep{kunsch:1989}. In
particular they note that although the block bootstrap accurately replicates
the first-order dependence structure of the original times series it fails to
reproduce second-order effects, because these are corrupted by the blocking
process. Use of an adjusted variance estimate to correct for the failure to
approximate second-order effects results, in turn, in an error rate of only
$O_{p}(T^{-2/3+\beta})$ for the block bootstrap. In contrast, the second-order
structure is shown to be preserved by the sieve. \cite{choi:hall:2000}
demonstrate that the performance of the sieve is robust to the selected order
for the autoregressive approximation, whilst noting that the choice of block
length and other tuning parameters can be crucial to the performance of the
block bootstrap. Moreover, as these authors also remark, the use of an
automated method such as Akaike's information criterion ($AIC$) to determine
the autoregressive order offers obvious practical advantages, again in
contrast with the situation that prevails for the block bootstrap, whereby
generic selection rules for the block length are unavailable. These
deficiences identified in the block bootstrap technique are likely to be
manifest with long range dependent data \textit{a-fortiori,} suggesting that
the sieve bootstrap is likely to be even more favoured for fractionally
integrated processes. For a review of block and sieve bootstrap methods and
further discussion of their associated features see \citet{politis:2003}.

We illustrate our proposed methods by means of a simulation study, in which we
examine the sieve bootstrap approximation to the sampling distribution of two
types of statistic that satisfy the relevant conditions for the convergence
results to hold. Firstly, we compare and contrast the performance of the raw
and the pre-filtered sieve bootstrap in correctly characterizing the known
finite sample properties of the sample mean under long memory. In particular,
we investigate the previously noted tendency of bootstrap techniques to
underestimate the true variance of the sample mean in this setting
\citep{hesterberg:1997}. The pre-filtering is shown to correct for the
distinct underestimation of the sampling variance still produced by the raw
sieve, the higher-order accuracy of the latter notwithstanding. Secondly, we
document the performance of the two bootstrap methods in estimating the
(unknown) sampling distributions of selected autocorrelation coefficients. We
undertake two exercises here. We begin by comparing the estimates of the
sampling distributions produced by the (raw) sieve bootstrap with those
produced via an Edgeworth approximation, in the region of the parameter space
where such an approximation is valid
\citep[see][]{lieberman:rousseau:zucker:2001}. The bootstrap method is shown
to produce distributions that are visually indistinguishable from those
produced by the \textrm{second-order} Edgeworth expansion which, in turn,
replicate the Monte Carlo estimates. Encouraged by the accuracy of the
bootstrap method in the case in which an analytical finite sample comparator
is available, we then proceed to assess the relative performance of the two
alternative sieve bootstrap methods - raw and pre-filtered - in the part of
the parameter space in which it is not. The pre-filtered method (in
particular) is shown to produce particularly accurate estimates of the
\textquotedblleft true\textquotedblright\ (Monte\ Carlo) distributions in this
region, augering well for its general usefulness in empirical settings.

The paper proceeds as follows. Section 2 briefly outlines the statistical
properties of autoregressive approximations to fractionally integrated
processes, and summarizes the properties of the raw sieve bootstrap in this
context. In Section 3 we present relevant Edgeworth expansions for a given
class of statistics, and exploit these representations to establish the stated
error rates for the raw sieve bootstrap technique. Section 4 outlines the
methodology underlying the pre-filtered sieve bootstrap and presents the
associated theory indicating the improvement obtained thereby. Details of the
simulation study are given in Section 5. Section 6 closes the paper with some
concluding remarks.

\section{Long memory processes, autoregressive approximation, and the sieve
bootstrap}

Let $y(t)$ for $t\in{\mathcal{Z}}$ denote a linearly regular,
covariance-stationary process with representation as in \eqref{Wold} where the
innovations and the impulse response coefficients satisfy the following conditions:

\begin{assumption}
\label{Ass1} The innovation process $\varepsilon(t)$ is ergodic and,
\begin{equation}
E\big[\varepsilon(t)\mid\mathcal{E}_{t-1}%
\big]=0~~~\mbox{and}~~~E\big[\varepsilon(t)^{2}\mid\mathcal{E}_{t-1}%
\big]=\sigma^{2}\,, \tag{ass1}%
\end{equation}
where $\mathcal{E}_{t}$ denotes the $\sigma$-algebra of events determined by
$\varepsilon(s)$, $s\leq t$. Furthermore, $E\big[\varepsilon(t)^{4}%
\big]<\infty$.
\end{assumption}

\begin{assumption}
\label{Ass2} The transfer function $k(z)=\sum_{j\geq0}k(j)z^{j}$ in the
representation of the process $y(t)$ is given by $k(z)=\kappa(z)/(1-z)^{d}$
where $|d|<0.5$ and $\kappa(z)$ satisfies $\kappa(z)\neq0$, $|z|\leq1$, and
$\sum_{j\geq0}j|\kappa(j)|<\infty$.
\end{assumption}

Assumption \ref{Ass1} imposes a classical martingale difference structure on
the innovations, the critical property of such a process that drives the
asymptotic results being that a martingale difference is uncorrelated with any
measurable function of its own past. Assumption \ref{Ass2} rules out the
possibility of a root at unity in $\kappa(z)$ canceling with $(1-z)^{d}$ and
implies that the underlying process admits an infinite-order autoregressive
($AR(\infty)$) representation. Assumptions 1 and 2 incorporate quite a wide
class of linear processes, including the popular ARFIMA family of models
introduced by \cite{granger:joyeux:1980} and \cite{hosking:1981}.

Under Assumptions \ref{Ass1} and \ref{Ass2} $y(t)=\bar{y}(t)+\varepsilon(t)$
where the {linear} predictor
\[
\bar{y}(t)=\sum_{j=1}^{\infty}\pi(j)y(t-j)\,,\quad\sum_{j=1}^{\infty}%
\pi(j)z^{j}=1-k(z)^{-1}\,,
\]
is the minimum mean squared error predictor (MMSEP) of $y(t)$ based on the
infinite past. The MMSEP of $y(t)$ based only on the finite past is then
\begin{equation}
\bar{y}_{h}(t)=\sum_{j=1}^{h}\pi_{h}(j)y(t-j)\equiv-\sum_{j=1}^{h}\phi
_{h}(j)y(t-j), \label{plinh}%
\end{equation}
where the minor reparameterization from $\pi_{h}$ to $\phi_{h}$ allows us, on
also defining $\phi_{h}(0)=1$, to conveniently write the corresponding
prediction error as
\begin{equation}
\varepsilon_{h}(t)=\sum_{j=0}^{h}\phi_{h}(j)y(t-j). \label{epsh}%
\end{equation}

The finite-order autoregressive coefficients $\phi_{h} (1),\ldots,\phi_{h}
(h)$ can be deduced from the Yule-Walker equations
\begin{equation}
\label{Y-W}\sum_{j=0}^{h} \phi_{h}(j)\gamma(j-k) =\delta_{0}(k) \sigma^{2}%
_{h}\,, \quad k=0,1,\ldots,h,
\end{equation}
in which $\gamma(\tau)=\gamma(-\tau)=E[y(t)y(t-\tau)]$, $\tau=0,1,\ldots$ is
the autocovariance function of the process $y(t)$, $\delta_{0}(k)$ is
Kronecker's delta (i.e., $\delta_{0}(k)=0\; \forall\; k\ne0;\; \delta_{0}(0)=1
$), and
\begin{equation}
\label{Varh}\sigma^{2}_{h} =E\big[\varepsilon_{h}(t)^{2}\big]
\end{equation}
is the prediction error variance associated with $\bar{y}_{h}(t)$.

The use of finite-order autoregressive models to approximate an unknown (but
suitably regular) process therefore requires that the optimal predictor
$\bar{y}_{h}(t)$ determined from the autoregressive model of order $h$
($AR(h)$) be a good approximation to the \textquotedblleft
infinite-order\textquotedblright\ predictor $\bar{y}(t)$ for sufficiently
large $h$. The asymptotic validity and properties of $AR(h)$ models when
$h\rightarrow\infty$ with the sample size $T$ under regularity conditions that
admit non-summable processes were established in \cite{poskitt:2007}. Briefly,
the order-$h$ prediction error $\varepsilon_{h}(t)$ converges to
$\varepsilon(t)$ in mean-square, the estimated sample-based covariances
converge to their population counterparts, though at a slower rate than for a
conventionally stationary process, and the least squares and Yule-Walker
estimators of the coefficients of the $AR(h)$ approximation are asymptotically
equivalent and consistent. Furthermore, order selection by $AIC$ is
asymptotically efficient in the sense of being equivalent to minimizing
Shibata's \citeyearpar{shibata:1980} figure of merit, discussed in more detail
in Section \ref{sim} in the context of the simulation experiment reported
therein. The sieve bootstrap, which works by \textquotedblleft
whitening\textquotedblright\ the data using an $AR(h)$ approximation, with the
dynamics of the process captured in the fitted autoregression, is accordingly
a plausible semi-parametric bootstrap technique for long memory processes.
Details of its application to fractional processes are given in
\cite{poskitt:2008}.

For convenience we present here the basic steps needed to generate a sieve
bootstrap realization of a process $y(t)$ (referred to as the sieve bootstrap
(SBS) algorithm hereafter):

\begin{enumerate}
\item[SB1.] \label{s1} Given data $y(t)$, $t=1,\ldots,T$, calculate the
parameter estimates of the $AR(h)$ approximation, denoted by $\bar{\phi}%
_{h}(1),\ldots,\bar{\phi}_{h}(h)$ and $\bar{\sigma}_{h}^{2}$, and evaluate the
residuals,
\[
\bar{\varepsilon}_{h}(t)=\sum_{j=0}^{h}\bar{\phi}_{h}(j)y(t-j)\,,\,t=1,\ldots
,T\,,
\]
using $y(1-j)=y(T-j+1)$, $j=1,\ldots,h$, as initial values. From
$\bar{\varepsilon}_{h}(t)$, $t=1,\ldots,T$, construct the standardized
residuals $\tilde{\varepsilon}_{h}(t)=(\bar{\varepsilon}_{h}(t)-\bar
{\varepsilon}_{h})/s_{\bar{\varepsilon}_{h}}$, where $\bar{\varepsilon}%
_{h}=T^{-1}\sum_{t=1}^{T}\bar{\varepsilon}_{h}(t)$ and $s_{\bar{\varepsilon
}_{h}}^{2}=T^{-1}\sum_{t=1}^{T}(\bar{\varepsilon}_{h}(t)-\bar{\varepsilon}%
_{h})^{2}$.

\item[SB2.] Let $\varepsilon^{+}_{h}(t)$, $t=1,\ldots,T$, denote a simple
random sample of \textit{i.i.d.} values drawn from
\[
U_{\bar{\varepsilon}_{h},T}(e)=T^{-1}\sum_{t=1}^{T}\mathbf{1}\{ \tilde
{\varepsilon}_{h}(t)\leq e\} \,,
\]
the probability distribution function that places a probability mass of $1/T$
at each of $\tilde{\varepsilon}_{h}(t)$, $t=1,\ldots,T$. Set $\varepsilon
^{*}_{h}(t)=\bar{\sigma}_{h}\varepsilon^{+}_{h}(t)$, $t=1,\ldots,T$.

\item[SB3.] Construct the sieve bootstrap realization $y^{*}(1),\ldots
,y^{*}(T)$ where $y^{*}(t)$ is generated from the autoregressive process
\[
\sum_{j=0}^{h}\bar{\phi}_{h}(j)y^{*}(t-j)=\varepsilon^{*}_{h}(t)\,,\,
t=1,\ldots,T\,,
\]
initiated at $y^{*}(1-j)=y(\tau-j+1)$, $j=1,\ldots,h$, where $\tau$ has the
discrete uniform distribution on the integers $h,\ldots,T$.
\end{enumerate}

Crucially, in the fractional case the rate of convergence of the coefficient
estimates $\bar{\phi}_{h}(1),\ldots,\bar{\phi}_{h}(h)$ evaluated in Step SB1
is dependent upon the value of the fractional index $d$.

\begin{theorem}
\label{consistentYW} Let $\sum_{j=0}^{h}\bar{\phi}_{h}(j)z^{j}$ denote the
Burg, least squares or Yule-Walker estimator of $\sum_{j=0}^{h}\phi
_{h}(j)z^{j}$. If $y(t)$ is a stationary process that satisfies Assumptions
\ref{Ass1} and \ref{Ass2} (given below) then for all $h\leq H_{T}=a(\log
T)^{c} $, $a>0$, $c<\infty$,
\[
\sum_{j=1}^{h}|\bar{\phi}_{h}(j)-\phi_{h}(j)|^{2}=O\left\{  h\left(
\frac{\log T}{T}\right)  ^{1-2d^{\prime}}\right\}  \quad\text{a.s.}%
\]
where $d^{\prime}=\max\{0,d\}$.
\end{theorem}

\noindent\textbf{Proof}: For the least squares and Yule-Walker estimators see
\citet[][Theorem 5 and Corollary 1]{poskitt:2007} and the associated
discussion. For the Burg estimator the result then follows from
\citet[][Theorem 1]{poskitt:1994}. $\hfill\qed$

\medskip Now consider a statistic $\mathbf{s}_{T}=(s_{1T},\ldots
,s_{mT})^{\prime}$, where $s_{iT}=s_{i}(y(1),\ldots,y(T))$ and each
$s_{i}(\cdot)$ for $i=1,\ldots,m$ is a suitably smooth function of the time
series values $y(1),\ldots,y(T)$. Let $F_{\mathbf{s}_{T}}(\mathbf{s})$ denote
the distribution function of $\mathbf{s}_{T}$ under $(\Omega,\mathfrak{F},P)$,
the original probability space. Let $\mathbf{s}_{T}^{\ast} $ be defined as for
$\mathbf{s}_{T}$ but with the observed realization replaced by $y^{\ast
}(1),\ldots,y^{\ast}(T)$, a realization obtained from the SBS algorithm, so
that $\mathbf{s}_{T}^{\ast}=(s_{1T}^{\ast},\ldots,s_{mT}^{\ast})^{\prime}$
where $s_{iT}^{\ast}=s_{i}(y^{\ast}(1),\ldots,y^{\ast}(T))$. Let
$F_{\mathbf{s}_{T}^{\ast}}(\mathbf{s})$ denote the distribution of
$\mathbf{s}_{T}^{\ast}$ under $(\Omega^{\ast},\mathfrak{F}^{\ast},P^{\ast})$,
the bootstrap probability space. As with $F_{\mathbf{s}_{T}}(\mathbf{s})$, the
analytical determination of $F_{\mathbf{s}_{T}^{\ast}}(\mathbf{s})$ is
generally intractable, but by simulating a large number, $B$, of independent
bootstrap realizations and calculating $\mathbf{s}_{T,b}^{\ast}$ for
$b=1,\ldots,B$, we can approximate $F_{\mathbf{s}_{T}^{\ast}}(\mathbf{s})$ by
the bootstrap empirical distribution function
\begin{equation}
\bar{F}_{\mathbf{s}_{T}^{\ast},B}(\mathbf{s})=B^{-1}\sum_{b=1}^{B}%
\mathbf{1}\{\mathbf{s}_{T,b}^{\ast}\leq\mathbf{s}\}\,. \label{bedf}%
\end{equation}
By the (strong) Glivenko-Cantelli Theorem
\[
\limsup_{B\rightarrow\infty}\sqrt{\frac{B}{2\log\log B}}\text{sup}%
_{\mathbf{s}}|\bar{F}_{\mathbf{s}_{T}^{\ast},B}(\mathbf{s})-F_{\mathbf{s}%
_{T}^{\ast}}(\mathbf{s})|\leq\frac{1}{2}\quad\text{a.s.}%
\]
and we can approximate $F_{\mathbf{s}_{T}^{\ast}}(\mathbf{s})$ arbitrarily
closely by taking the number of bootstrap realizations sufficiently large. The
idea behind the bootstrap is that the distribution of $\mathbf{s}_{T}^{\ast}$
under $(\Omega^{\ast},\mathfrak{F}^{\ast},P^{\ast})$ should mimic that of
$\mathbf{s}_{T}$ under $(\Omega,\mathfrak{F},P)$ and we can therefore
anticipate that $\bar{F}_{\mathbf{s}_{T}^{\ast},B}(\mathbf{s})$ will also
approximate $F_{\mathbf{s}_{T}}(\mathbf{s})$ closely provided $F_{\mathbf{s}%
_{T}^{\ast}}(\mathbf{s})$ is sufficiently near to $F_{\mathbf{s}_{T}%
}(\mathbf{s})$.

That the autoregressive sieve bootstrap provides a valid approximation to
$F_{\mathbf{s}_{T}}(\mathbf{s})$ under the current assumptions can be
established by generalizing the arguments of
\cite{kreiss:paparoditis:politis:2011} using the extension of Baxter's
inequality due to \cite{inoue:kasahara:2006}. It can be shown
\citep{poskitt:2008} that for the class of linear statistics considered in
\citet[][Section
2.1]{kunsch:1989} and \citet[][Section 3.3]{buhlmann:1997} we have
$\eta(F_{\mathbf{s}_{T}^{\ast}},F_{\mathbf{s}_{T}})=o(T^{-1/2(1-2d^{\prime
})+\beta})$ for all $\beta>0$, wherein $d^{\prime}=\max\{0,d\}$ and
$\eta(F_{X},F_{Y})$ denotes Mallow's measure of the distance between two
probability distributions $F_{X}$ and $F_{Y}$. Mallows metric is equivalent to
weak convergence \citep[][Lemma
8.3]{bickel:freedman:1981} and in conjunction with a convergence rate of
$T^{-1/2(1-2d^{\prime})+\beta}$ this intimates that use of the sieve bootstrap
may be little better than applying a central limit approximation. However, in
what follows we show that for a more restricted range of statistics (albeit
one that intersects with the linear class) the convergence rate can be
improved upon, and that the rate established by \cite{choi:hall:2000} in the
short memory case can in fact be generalized to long memory processes.

\section{Higher Order Improvements for the Sieve Bootstrap\label{higher}}

Let us suppose that $F_{\mathbf{s}_{T}}(\mathbf{s})$ is absolutely continuous
with respect to Lebesgue measure, differentiable for all $\mathbf{s}$, and
that the following assumptions are satisfied.

\begin{assumption}
\label{linStat} There exists a function $M_{T}$ (possibly stochastic) and a
constant $M<\infty$ such that
\[
\Vert\mathbf{s}_{T}^{\ast}-\mathbf{s}_{T}\Vert^{2}\leq mM_{T}T^{-1}\sum
_{t=1}^{T}(y^{\ast}(t)-y(t))^{2},
\]
where $M_{T}$ is bounded (in probability) by $M$.
\end{assumption}

Whilst defining a more restrictive class (overall) than the linear class, it
remains the case that a broad range of statistics used in the analysis of time
series satisfy Assumption \ref{linStat}, see \citep[][Lemma
1]{poskitt:2008}. As highlighted in the latter, this set includes the sample
mean and the sample autocovariances, autocorrelations and partial
autocorrelations. Further examples include the discrete Fourier transform and
OLS regression coefficients. The former follows on setting
\[
\mathbf{s}_{T}^{\ast}-\mathbf{s}_{T}=\frac{1}{(2\pi T)^{\half}}\sum_{t=1}%
^{T}\left[
\begin{array}
[c]{c}%
y^{\ast}(t)-y(t)\\
(y^{\ast}(t)-y(t))\exp(-\imath2\pi t/T)\\
\vdots\\
(y^{\ast}(t)-y(t))\exp(-\imath2\pi(T-1)t/T)\\
\end{array}
\right]  \,,
\]
with the validity of Assumption \ref{linStat}, using $M=1/(2\pi)$, now a
direct consequence of Parseval's theorem. For the latter, let $\bx(t)=(x_{1}%
(t),\ldots,x_{k}(t))^{\prime}$ denote a vector of regressors that satisfy
$\liminf_{T\rightarrow\infty}\lambda_{\min}[T^{-1}\sum_{t=1}^{T}%
\bx(t)\bx(t)^{\prime}]\geq\lambda>0$ and set $\bk(t)=(\sum_{t=1}%
^{T}\bx(t)\bx(t)^{\prime})^{-1}\bx(t)$. Then
\[
\Vert\mathbf{s}_{T}^{\ast}-\mathbf{s}_{T}\Vert^{2}=\Vert\sum_{t=1}%
^{T}\bk(t)(y^{\ast}(t)-y(t))\Vert^{2}\leq\sum_{t=1}^{T}\Vert\bk(t)\Vert
^{2}\sum_{t=1}^{T}(y^{\ast}(t)-y(t))^{2}%
\]
and Assumption \ref{linStat} holds because $\sum_{t=1}^{T}\Vert\bk(t)\Vert
^{2}\leq k/\lambda T$. As a point of interest, applying the immediately
preceding regression inequality to the log-periodogram regression estimator of
$d$ \citep{geweke:porter:1983} and using the inequalities $\log(1+x)\leq x$,
$x\geq0$, and $||z|-|z^{\ast}||\leq|z-z^{\ast}|$ for any pair of complex
numbers $z$ and $z^{\ast}$, we also find that the log-periodogram regression
estimator satisfies Assumption \ref{linStat} on application of the bound on
the discrete Fourier transform.

The following assumption implicitly characterizes moment conditions under
which a valid Edgeworth expansion exists for statistics in the class described
by Assumption \ref{linStat}.

\begin{assumption}
\label{chfAss} Let $\psi_{T}(\btau)=E[\exp(\imath\btau^{\prime}\mathbf{s}%
_{T})]$ denote the characteristic function of $\mathbf{s}_{T}$ where
$\btau=(\tau_{1},\ldots,\tau_{m})^{\prime}$ and let $\partial^{j}\log\psi
_{T}(\btau)/\partial\btau^{j}$ denote the vector of $j$th-order partial
derivatives corresponding to $\partial^{j}\log\psi_{T}(\btau)/\partial\tau
_{1}^{j_{1}}\cdots\partial\tau_{m}^{j_{m}}$ for all non-negative integers
$j_{1},\ldots,j_{m}$ satisfying $\sum_{l=1}^{m}j_{l}=j$. Then firstly, for any
$\delta>0$ and some integer $r\geq3$ the conditions
\[
\int_{\Vert\btau\Vert>\delta\sqrt{T}}|\psi_{T}(\btau)|^{2}d\btau=o(T^{2-r}%
)~\mbox{and}\newline\int_{\Vert\btau\Vert>\delta\sqrt{T}}\left\vert
\frac{\partial^{s}\psi_{T}(\btau)}{\partial\tau_{l}^{s}}\right\vert
^{2}d\btau=O(T^{1-r}),~l=1,\ldots,m,
\]
hold where $s=[m/2]+1$. Secondly, $\partial^{q}\log\psi_{T}(\btau)/\partial
\btau^{q}$ exists for all $\btau$ in a neighbourhood of the origin and
$\lim_{\Vert\btau\Vert\rightarrow0}T^{-1}\partial^{q}\log\psi_{T}%
(\btau)/\partial\btau^{q}$ exists as $T\rightarrow\infty$ for all
$q=1,\ldots,q^{\prime}=\max\{s,r+1\}$.
\end{assumption}

Here $E$ denotes expectation taken with respect to the probability measure
induced by the original probability space $(\Omega,\mathfrak{F},P)$.
Assumption \ref{chfAss} summarizes Assumptions 1 and 2 of
\cite{taniguchi:1984}, which in turn encompass Assumptions 2 through 4 of
\cite{durbin:1980}, to which we refer for an in depth discussion. In any
particular instance, satisfaction of the conditions in Assumption \ref{chfAss}
must be ascertained and may occur only in particular parts of the parameter
space, such as in the case of the sample autocorrelation function investigated
in Section \ref{auto} (See the Appendix for details).

Let $\mathbf{V}_{T}=T^{-1}E\left[  (\mathbf{s}_{T}-E[\mathbf{s}_{T}%
])(\mathbf{s}_{T}-E[\mathbf{s}_{T}])^{\prime}\right]  $ and set $\bzeta_{T}%
=\mathbf{V}_{T}^{-1/2}T^{-\half}(\mathbf{s}_{T}-E[\mathbf{s}_{T}])$. If we
suppose that $\mathbf{V}_{T}=\mathbf{V}+o(1)$ where $\mathbf{V}$ is positive
definite, then Assumption \ref{chfAss} ensures the validity of the Edgeworth
approximation
\begin{equation}
\textmd{P}(\bzeta_{T}\leq\mathbf{z})=G(\mathbf{z})+\sum_{j=3}^{r}%
T^{1-j/2}p_{j}(\mathbf{z},\mathbf{K}_{r})g(\mathbf{z})+o(T^{1-r/2})
\label{Edge1}%
\end{equation}
uniformly in $\mathbf{z}$, where $G(\mathbf{z})$ denotes the distribution
function of a Gaussian $\mathbb{N}(\mathbf{0},\mathbf{I}_{m})$ random vector,
$g(\mathbf{z})$ the corresponding density, and $p_{j}(\mathbf{z}%
,\mathbf{K}_{r})$ is a polynomial function of degree $j$ in $\mathbf{z}$ whose
coefficients are polynomials in the elements of the cumulants $\mathbf{K}%
_{r}=(\mathbf{k}_{1}^{\prime},\ldots,\mathbf{k}_{r}^{\prime})^{\prime}$,
$\mathbf{k}_{r}=\imath^{-r}\partial^{r}\log\psi_{T}(\mathbf{0})/\partial
\boldsymbol{\tau}^{r}$. See Theorem 1 of \cite{taniguchi:1984} and
\cite{durbin:1980}.\footnote{We have thus far supposed that $\mathbf{s}_{T}$
is a continuous random variable. For extension to the lattice case see
\citet[][\S 5.4]{durbin:1980}}

Similarly, if $E^{\ast}$ denotes expectation taken with respect to the
probability space $(\Omega^{\ast},\mathfrak{F}^{\ast},P^{\ast})$ and
$\bzeta_{T}^{\ast}=\mathbf{V}_{T}^{\ast-1/2}T^{-\half}(\mathbf{s}_{T}^{\ast
}-E^{\ast}[\mathbf{s}_{T}^{\ast}])$ where $\mathbf{V}_{T}^{\ast}=T^{-1}%
E^{\ast}\left[  (\mathbf{s}_{T}^{\ast}-E^{\ast}[\mathbf{s}_{T}^{\ast
}])(\mathbf{s}_{T}^{\ast}-E^{\ast}[\mathbf{s}_{T}^{\ast}])^{\prime}\right]  $,
then under appropriate regularity
\begin{equation}
\textmd{P}^{\ast}(\bzeta_{T}^{\ast}\leq\mathbf{z})=G(\mathbf{z})+\sum
_{j=3}^{r}T^{1-j/2}p_{j}(\mathbf{z},\mathbf{K}_{r}^{\ast})g(\mathbf{z}%
)+o(T^{1-r/2}), \label{Edge2}%
\end{equation}
where $\mathbf{K}_{r}^{\ast}=(\mathbf{k}_{1}^{\ast\prime},\ldots
,\mathbf{k}_{r}^{\ast\prime})^{\prime}$, $\mathbf{k}_{r}^{\ast}=\imath
^{-r}\{\partial^{r}\log\psi_{T}^{\ast}(\mathbf{0})/\partial\boldsymbol{\tau
}^{r}\}$, $\psi_{T}^{\ast}(\boldsymbol{\tau})=E^{\ast}[\exp(\imath
\boldsymbol{\tau}^{\prime}\mathbf{s}_{T}^{\ast})]$.

A comparison of \eqref{Edge1} and \eqref{Edge2} for $r\geq4$ now indicates
that
\begin{equation}
\sup_{\mathbf{z}}|\textmd{P}^{\ast}(\bzeta_{T}^{\ast}\leq\mathbf{z}%
)-\textmd{P}(\bzeta_{T}\leq\mathbf{z})|=T^{-1/2}O(\Vert\mathbf{K}_{r}^{\ast
}-\mathbf{K}_{r}\Vert)+o(T^{-1})\,. \label{Edge3}%
\end{equation}
Noting that $\textmd{P}^{\ast}$ depends on $\textmd{P}$ so the elements of
$\mathbf{K}_{r}^{\ast}$, which are constants relative to $\textmd{P}^{\ast}$,
are random variables relative to $\textmd{P}$, we see that if $\Vert
\mathbf{K}_{4}^{\ast}-\mathbf{K}_{4}\Vert=o_{p}(T^{-1/2}\varrho_{T}) $ then
\eqref{Edge3} implies that the bootstrap probability will have an error rate
of $O_{p}(T^{-1}\varrho_{T})$. In their investigation of coverage accuracy
\citet[Appendix A.2]{choi:hall:2000} used this type of argument when analyzing
the subset of linear statistics characterized by polynomial products; and it
was also employed by \cite{andrews:lieberman:marmer:2006} to show that the
parametric bootstrap based on the (approximate) MLE of parameters in a
Gaussian long memory model achieves an error rate of order $T^{-1}\log T$ for
a one sided confidence interval. Using this approach we can establish
analogous results for the sieve bootstrap in the long memory case, and for the
class of statistics encompassed by Assumption \ref{linStat}.

\begin{theorem}
\label{BT} Suppose that the statistic $\mathbf{s}_{T}$ satisfies Assumption
\ref{linStat} and Assumption \ref{chfAss} with $r\geq4$ when calculated from a
process $y(t)$ that also satisfies Assumptions \ref{Ass1} and \ref{Ass2} .
Then for all $\beta>0$
\[
\sup_{\mathbf{z}}|\textmd{P}^{\ast}(\bzeta_{T}^{\ast}\leq\mathbf{z}%
)-\textmd{P}(\bzeta_{T}\leq\mathbf{z})|=O_{p}(T^{-(1-d^{\prime})+\beta})\,.
\]

\end{theorem}

The proof of Theorem \ref{BT} relies on the following lemma. The heuristics
behind the proof are straightforward; convergence of Mallow's metric implies
convergence in distribution and hence, via the Cram\'{e}r-Levy continuity
theorem, convergence of the characteristic function and the associated moments
and cumulants \citep[See Lemma 8.3 of][]{bickel:freedman:1981}.

\begin{lemma}
\label{mbound} \label{characf} Suppose that the process $y(t)$ satisfies
Assumptions \ref{Ass1} and \ref{Ass2}, and that the statistic $\mathbf{s}_{T}
$ satisfies Assumption \ref{linStat}. Then $E[E^{\ast}[\Vert\mathbf{s}%
_{T}^{\ast}-\mathbf{s}_{T}\Vert^{2}]]=o(T^{-(1-2d^{\prime})+\beta})$ for all
$\beta>0$. Moreover, for any $\epsilon>0$ and $\eta> 0$, no matter how small,
there exists a $T_{\epsilon,\eta}$ such that
\[
\textmd{P}\left(  T^{-1/2(1-2d^{\prime})+\beta}|\psi_{T}^{\ast}(\btau)-\psi
_{T}(\btau)|<\epsilon\right)  >1-\eta
\]
for all $T>T_{\epsilon,\eta}$ uniformly in $\btau$, $\Vert\btau\Vert\leq
T^{\beta/2}$.
\end{lemma}

\noindent\textbf{Proof}: Given that $\mathbf{s}_{T}$ satisfies Assumption
\ref{linStat}, it follows that $\Vert\mathbf{s}_{T}^{\ast}-\mathbf{s}_{T}%
\Vert^{2}\leq mMT^{-1}\sum_{t=1}^{T}(y^{\ast}(t)-y(t))^{2}$ where $M<\infty$.
Arguing as in the proof of Theorem 2 of \citep[][p.246-248]{poskitt:2008} we
therefore have $E[E^{\ast}[\Vert\mathbf{s}_{T}^{\ast}-\mathbf{s}_{T}\Vert
^{2}]]\leq mME[E^{\ast}[(y^{\ast}(t)-y(t))^{2}]]=o(T^{-(1-2d^{\prime})+\beta
})$ for all $\beta>0$, which yields the first part of the lemma.

To prove the second part of the lemma note that since $\exp(\imath x)$ is
continuous with a continuous and uniformly bounded derivative it satisfies a
Lipschitz condition. Thus, for all $\boldsymbol{\tau}$ such that
$\Vert\boldsymbol{\tau}\Vert\leq T^{\beta/2}$ there exists a $K<\infty$ such
that $|\exp(\imath\boldsymbol{\tau}^{\prime}\mathbf{x})-\exp(\imath
\boldsymbol{\tau}^{\prime}\mathbf{y})|\leq KT^{\beta/2}\Vert\mathbf{x}%
-\mathbf{y}\Vert$. Then, as in \citet[][p. 1212]{bickel:freedman:1981},
\[
E[|\psi_{T}^{\ast}(\boldsymbol{\tau})-\psi_{T}(\boldsymbol{\tau})|]\leq
E[E^{\ast}[|\exp(\imath\boldsymbol{\tau}\mathbf{s}_{T}^{\ast})-\exp
(\imath\boldsymbol{\tau}\mathbf{s}_{T})|]]\leq KT^{\beta/2}E[E^{\ast}%
[\Vert\mathbf{s}_{T}^{\ast}-\mathbf{s}_{T}\Vert]]\,.
\]
But $E[E^{\ast}[\Vert\mathbf{s}_{T}^{\ast}-\mathbf{s}_{T}\Vert]]\leq
E[E^{\ast}[\Vert\mathbf{s}_{T}^{\ast}-\mathbf{s}_{T}\Vert^{2}]]^{1/2}%
=o(T^{-1/2(1-2d^{\prime}-\beta)})$. Application of Markov's inequality
completes the proof. $\hfill\qed$

\begin{corollary}
\label{cumul} Suppose that the process $y(t)$ satisfies Assumptions \ref{Ass1}
and \ref{Ass2}, and that the statistic $\mathbf{s}_{T}$ satisfies Assumptions
\ref{linStat} and \ref{chfAss}. Then for all $\beta>0$ we have $\Vert
\mathbf{K}_{q}^{\ast}-\mathbf{K}_{q}\Vert=o_{p}(T^{-1/2+d^{\prime}+\beta})$,
$q=1,\ldots,q^{\prime}=\max\{[m/2]+1,r+1\}$.
\end{corollary}

\noindent\textbf{Proof}: Using the expression $\log\psi_{T}^{\ast
}(\boldsymbol{\tau})-\log\psi_{T}(\boldsymbol{\tau})=\log\left(  1+(\psi
_{T}^{\ast}(\boldsymbol{\tau})-\psi_{T}(\boldsymbol{\tau}))/\psi
_{T}(\boldsymbol{\tau})\right)  $ and the fact that $\log(1+x)=x+O(|x|^{2})$
for $x$ in a neighbourhood of the origin we have $\log\psi_{T}^{\ast
}(\boldsymbol{\tau})-\log\psi_{T}(\boldsymbol{\tau})=(\psi_{T}^{\ast
}(\boldsymbol{\tau})-\psi_{T}(\boldsymbol{\tau}))/\psi_{T}(\boldsymbol{\tau
})+O\left(  \left\vert (\psi_{T}^{\ast}(\boldsymbol{\tau})-\psi_{T}%
(\boldsymbol{\tau}))/\psi_{T}(\boldsymbol{\tau})\right\vert ^{2}\right)
=o_{p}(T^{-1/2(1-2d^{\prime})+\beta})$ uniformly in $\boldsymbol{\tau}$ by
Lemma \ref{characf}.

Now set
\begin{align*}
\varphi_{T}^{\ast}(\mathbf{t};\boldsymbol{\tau})=  &  \frac{\log\psi_{T}%
^{\ast}(\mathbf{t})-\log\psi_{T}^{\ast}(\boldsymbol{\tau})}{\Vert
\mathbf{t}-\boldsymbol{\tau}\Vert}-\left(  \frac{\partial\log\psi
_{T}(\boldsymbol{\tau})}{\partial\boldsymbol{\tau}}\right)  ^{\prime}%
\frac{\mathbf{t}-\boldsymbol{\tau}}{\Vert\mathbf{t}-\boldsymbol{\tau}\Vert
}\quad\text{and}\\
\Delta_{T}^{\ast}(\mathbf{t};\boldsymbol{\tau})=  &  \frac{\log\psi_{T}^{\ast
}(\mathbf{t})-\log\psi_{T}^{\ast}(\boldsymbol{\tau})}{\Vert\mathbf{t}%
-\boldsymbol{\tau}\Vert}-\frac{\log\psi_{T}(\mathbf{t})-\log\psi
_{T}(\boldsymbol{\tau})}{\Vert\mathbf{t}-\boldsymbol{\tau}\Vert}\,,
\end{align*}
for $\mathbf{t}\neq\boldsymbol{\tau}$, and let $\epsilon>0$ be given. Then
\begin{align*}
|\varphi_{T}^{\ast}(\mathbf{t};\boldsymbol{\tau})|\leq &  \left\vert
\Delta_{T}^{\ast}(\mathbf{t};\boldsymbol{\tau})\right\vert \\
&  +\left\vert \frac{\log\psi_{T}(\mathbf{t})-\log\psi_{T}(\boldsymbol{\tau}%
)}{\Vert\mathbf{t}-\boldsymbol{\tau}\Vert}-\left(  \frac{\partial\log\psi
_{T}(\boldsymbol{\tau})}{\partial\boldsymbol{\tau}}\right)  ^{\prime}%
\frac{\mathbf{t}-\boldsymbol{\tau}}{\Vert\mathbf{t}-\boldsymbol{\tau}\Vert
}\right\vert
\end{align*}
and by definition of the differential \citep[][Section 6.4]{apostol:1960}
\[
\lim_{\Vert\mathbf{t}-\boldsymbol{\tau}\Vert\rightarrow0}|\varphi_{T}^{\ast
}(\mathbf{t};\boldsymbol{\tau})|\leq\lim_{\Vert\mathbf{t}-\boldsymbol{\tau
}\Vert\rightarrow0}\left\vert \Delta_{T}^{\ast}(\mathbf{t};\boldsymbol{\tau
})\right\vert +\epsilon\,.
\]
Since $\log\psi_{T}^{\ast}(\boldsymbol{\tau})-\log\psi_{T}(\boldsymbol{\tau
})=o_{p}(T^{-1/2(1-2d^{\prime})+\beta})$ uniformly in $\boldsymbol{\tau}$ we
can interchange limiting operations \citep[][Theorem
13.3]{apostol:1960} to give
\begin{align*}
\lim_{T\rightarrow\infty}\lim_{\Vert\mathbf{t}-\boldsymbol{\tau}%
\Vert\rightarrow0}\left\vert \Delta_{T}^{\ast}(\mathbf{t};\boldsymbol{\tau
})\right\vert \leq &  \lim_{\Vert\mathbf{t}-\boldsymbol{\tau}\Vert
\rightarrow0}\lim_{T\rightarrow\infty}\frac{|\log\psi_{T}^{\ast}%
(\mathbf{t})-\log\psi_{T}(\mathbf{t})|+|\log\psi_{T}^{\ast}(\boldsymbol{\tau
})-\log\psi_{T}(\boldsymbol{\tau})|}{\Vert\mathbf{t}-\boldsymbol{\tau}\Vert}\\
&  =o_{p}(T^{-1/2(1-2d^{\prime})+\beta})\,.
\end{align*}
Hence we can conclude that for all $T$ sufficiently large $\lim_{\Vert
\mathbf{t}-\boldsymbol{\tau}\Vert\rightarrow0}|\varphi_{T}^{\ast}%
(\mathbf{t};\boldsymbol{\tau})|\leq2\epsilon$ and $\log\psi_{T}^{\ast
}(\boldsymbol{\tau})$ has a differential at $\boldsymbol{\tau}$, since
$\epsilon$ is arbitrary, and
\[
\lim_{T\rightarrow\infty}\left\vert \frac{\partial\log\psi_{T}^{\ast
}(\boldsymbol{\tau})}{\partial\tau_{j}}-\frac{\partial\log\psi_{T}%
(\boldsymbol{\tau})}{\partial\tau_{j}}\right\vert \leq\lim_{h\rightarrow0}%
\lim_{T\rightarrow\infty}\left\vert \Delta_{T}^{\ast}(\boldsymbol{\tau
}+h\mathbf{u}_{j};\boldsymbol{\tau})\right\vert =o_{p}(T^{-1/2(1-2d^{\prime
})+\beta}),
\]
where $\mathbf{u}_{j}=(0,\ldots,0,1,0\ldots,0)^{\prime}$, the $j$th unit
vector, the existence of the gradient vector $\partial\log\psi_{T}^{\ast
}(\boldsymbol{\tau})/\partial\boldsymbol{\tau}$ being part of the conclusion
\citep[See][Theorem
6.13]{apostol:1960}. Thus, by definition of the first-order cumulant, we have
$\Vert\mathbf{K}_{1}^{\ast}-\mathbf{K}_{1}\Vert=o_{p}(T^{-1/2+d^{\prime}%
+\beta})$.

A parallel argument, with $\log\psi_{T}^{\ast}(\cdot)$ and $\log\psi_{T}%
(\cdot)$ replaced by $\partial^{j}\log\psi_{T}^{\ast}(\cdot)/\partial
\boldsymbol{\tau}^{j}$ and $\partial^{j}\log\psi_{T}(\cdot)/\partial
\boldsymbol{\tau}^{j}$, respectively, and $\partial\log\psi_{T}(\cdot
)/\partial\boldsymbol{\tau}$ replaced by $\partial^{j+1}\log\psi_{T}%
(\cdot)/\partial\boldsymbol{\tau}^{j+1}$, shows that $\partial^{j+1}\log
\psi_{T}^{\ast}(\cdot)/\partial\boldsymbol{\tau}^{j+1}$ exists and
$\Vert\mathbf{K}_{j+1}^{\ast}-\mathbf{K}_{j+1}\Vert=o_{p}(T^{-1/2+d^{\prime
}+\beta})$. Induction on $j=1,\ldots,q^{\prime}$ completes the proof.
$\hfill\qed$

\medskip

\noindent\textbf{Proof of Theorem \ref{BT}}: By construction the bootstrap
innovations $\varepsilon_{h}^{\ast}(t)$ in Step SB2 of the sieve bootstrap
satisfy Assumption \ref{Ass1}, and the sieve bootstrap process $y^{\ast}(t)$
produced in Step SB3 satisfies Assumption \ref{Ass2}. By definition, the
statistics $\mathbf{s}_{T}$ and $\mathbf{s}_{T}^{\ast}$ satisfy Assumption
\ref{linStat} and Assumption \ref{chfAss} with $r\geq4$, and Assumption
\ref{chfAss} validates the formal Edgeworth expansions in \eqref{Edge2} and
\eqref{Edge1}. Corollary \ref{cumul} (using Lemma \ref{characf}) implies that
$\Vert\mathbf{K}_{4}^{\ast}-\mathbf{K}_{4}\Vert=o_{p}(T^{-1/2}\varrho_{T})$
where $\varrho_{T}=o(T^{d^{\prime}+\beta})$ and Theorem \ref{BT} then follows
from equation \eqref{Edge3}.$\hfill\qed$

\medskip

Theorem \ref{BT} indicates the refinements that are possible using the sieve
bootstrap. For example, $S(q)=\{\mathbf{z}:\mathbf{z}^{\prime}\mathbf{z}\leq
q\}$ is a compact, Borel--measurable set in $\mathbb{R}^{m}$ that has finite
probability measure with respect to both $P$ and $P^{\ast}$. Now let
$q_{\alpha}^{\ast}$ be such that the Lebesgue--Stieltjes integral satifies the
following equality,
\[
\int_{S(q_{\alpha}^{\ast})}d\textmd{P}^{\ast}(\bzeta_{T}^{\ast}\leq
\mathbf{z})=1-\alpha\,.
\]
Then $S(q_{\alpha}^{\ast})$ is a raw sieve bootstrap $(1-\alpha)100\%$
elliptical percentile set for $\mathbf{s}_{T}$. Now, from Theorem \ref{BT} it
follows that
\begin{align*}
|(1-\alpha)-\int_{S(q_{\alpha}^{\ast})}d\textmd{P}(\bzeta_{T}\leq\mathbf{z})|
&  \leq\int_{S(q_{\alpha}^{\ast})}|d\textmd{P}^{\ast}(\bzeta_{T}^{\ast}%
\leq\mathbf{z})-d\textmd{P}(\bzeta_{T}\leq\mathbf{z})|\\
&  =\frac{(\pi q_{\alpha}^{\ast})^{m/2}}{\Gamma(m/2+1)}\,O_{p}%
(T^{-(1-d^{\prime})+\beta})
\end{align*}
for all $\beta>0$. This leads to a coverage probability for $S(q_{\alpha
}^{\ast})$ of $(1-\alpha)+O_{p}(T^{-(1-d)+\beta})$ for all $\beta>0$ when
$d\in(0,0.5)$, the long memory case, compared to $(1-\alpha)+O_{p}%
(T^{-1+\beta})$ when $d\in(-0.5,0]$, the short memory and anti-persistent
cases \citep[\textit{cf}.][]{choi:hall:2000}. Calibration of the percentile
sets using the double-bootstrap may be possible, but we will not pursue this
here. We will, however, investigate in the following section an adaptation of
the sieve bootstrap that improves the convergence rate by removing the
dependence on the fractional index $d$.

\section{The Pre-Filtered Sieve Bootstrap\label{pfsbs}}

Theorem \ref{consistentYW} indicates that the convergence of $\sum_{j=0}%
^{h}\bar{\phi}_{h}(j)z^{j}$ to $\sum_{j=0}^{h}\phi_{h}(j)z^{j}$ is slower the
larger is the value of $d$, and Theorem \ref{BT} shows that this feature is
passed on to the raw sieve bootstrap itself (and the associated coverage
probabilities of sets). Specifically, the closer is $d$ to zero the closer the
convergence rate will be to the rate achieved with short memory and
anti-persistent processes, namely $O_{p}(T^{-1+\beta}).$ Given the empirical
regularity of estimated values of $d$ in the range $(0,0.5)$, calculating a
preliminary estimate of $d$ and constructing a filtered version of the data to
which the AR approximation and sieve bootstrap are applied before inverse
filtering, may therefore yield advantages in terms of convergence.

With this in mind, let us suppose that a preliminary estimate $\widehat{d}$ of
$d$ is available such that $\widehat{d}-d\in N_{\delta}=\{x:|x|<\delta\} $
where $0<\delta<0.5$. For any $d>-1$ let $\alpha_{j}^{(d)}$, $j=0,1,2,\ldots$,
denote the coefficients of the fractional difference operator when expressed
in terms of its binomial expansion,
\begin{align*}
(1-z)^{d}=\sum_{j=0}^{\infty}\alpha_{j}^{(d)}z^{j}  &  =1+\sum_{j=1}^{\infty
}\left(  \frac{\Gamma(j-d)}{\Gamma(-d)\Gamma(j+1)}\right)  z^{j}\\
&  =1+\sum_{j=1}^{\infty}\left(  \prod_{0<k\leq j}\frac{k-1-d}{k}\right)
z^{j}\,,
\end{align*}
and set
\[
w(t)=\sum_{j=0}^{t-1}\alpha_{j}^{(d)}y(t-j)\,,\quad t=1,\ldots,T\,.
\]
Using the preliminary estimate $\widehat{d}$, pre-filtered sieve bootstrap
realizations of $y(t)$ can now be generated as follows:

\begin{enumerate}
\item[PFSBS1.] Calculate the coefficients of the filter $(1-z)^{\widehat{d}}$
and from the data generate the filtered values
\[
\widehat{w}(t)=\sum_{j=0}^{t-1}\alpha_{j}^{(\widehat{d})}y(t-j)
\]
for $t=1,\ldots,T$.

\item[PFSBS2.] Fit an AR approximation to $\widehat{w}(t)$ and generate a
sieve bootstrap sample $\widehat{w}^{\ast}(t)$, $t=1,\ldots,T$, of the
filtered data as in Steps SB1--SB3 of the SBS algorithm.

\item[PFSBS3.] Using the coefficients of the (inverse) filter
$(1-z)^{-\widehat{d}}$ construct a corresponding pre--filtered sieve bootstrap
draw
\begin{equation}
\widehat{y}^{\ast}(t)=\sum_{j=0}^{t-1}\alpha_{j}^{(-\widehat{d})}%
\widehat{w}^{\ast}(t-j) \label{bdraw}%
\end{equation}
of $y(t)$ for $t=1,\ldots,T$.
\end{enumerate}

We will refer to this as the PFSBS algorithm.

Note that the process
\[
(1-z)^{\widehat{d}}y(t)=\frac{\kappa(z)}{(1-z)^{d-\widehat{d}}}\,\varepsilon
(t)
\]
has fractional index $d-\widehat{d}$. By assumption $|d-\widehat{d}|<\delta$
and the error in the AR approximation fitted in Step PFSBS2 will accordingly
be of order $O(h\left(  \log T/T\right)  ^{1-2\delta})$ or smaller (Theorem
\ref{consistentYW}). That this level of accuracy is transferred to the
pre--filtered sieve bootstrap realizations $\widehat{y}^{\ast}(t)$ of $y(t)$,
via the sieve bootstrap draws $\widehat{w}^{\ast}(t) $ of $\widehat{w}(t)$,
and hence to the pre-filtered sieve bootstrap approximation to the sampling
distribution of the statistic $\mathbf{s}_{T}$, rests upon the following proposition.

\begin{proposition}
\label{prefil} Suppose that the process $y(t)$ satisfies Assumptions
\ref{Ass1} and \ref{Ass2}. Let $\widehat{d}$ be such that $\widehat{d}%
\in(-0.5,0.5(1-\epsilon))$ for some $\epsilon>0$. Then there exists a constant
$G<\infty$, independent of $\widehat{d}$, such that $E[E^{*}[(\widehat{y}%
^{*}(t)-y(t))^{2}]]\leq GE[E^{*}[(\widehat{w}^{*}(t)-\widehat{w}(t))^{2}]]$.
\end{proposition}

\noindent\textbf{Proof}: By construction
\[
y(t)=\sum_{j=0}^{t-1}\alpha^{(-\widehat{d})}_{j}\widehat{w}(t-j)\,,
\]
and subtracting on the left and right hand sides in \eqref{bdraw} it follows
that
\begin{equation}
\label{wdif}\widehat{y}^{*}(t)-y(t) = \sum_{j=0}^{t-1}\alpha^{(-\widehat{d}%
)}_{j}\{ \widehat{w}^{*}(t-j)-\widehat{w}(t-j)\}\quad t=1,\ldots,T
\end{equation}
for all possible pairs $(y(t),\widehat{y}^{*}(t))$, $t=1,\ldots,T$, in the
product space generated by $(\Omega\otimes\Omega^{*},\mathfrak{F}%
\otimes\mathfrak{F}^{*},P(P^{*}))$ with joint distribution corresponding to
the marginal and conditional probability measures $P$ and $P^{*}$.

Now let $Z_{\{\widehat{w}^{\ast}-\widehat{w}\}}^{T}(\lambda)$, $\lambda
\in\lbrack0,2\pi]$, denote the finite sample spectral measure associated with
the process $\widehat{w}^{\ast}(t)-\widehat{w}(t)$, $t=1,\ldots,T$, which we
define to be%
\[
Z_{\{\widehat{w}^{\ast}-\widehat{w}\}}^{T}(\lambda)=\frac{1}{2\pi}\sum
_{t=1}^{T}\frac{e^{-\imath\lambda t}-1}{\imath t}\{\widehat{w}^{\ast
}(t)-\widehat{w}(t)\}=\frac{1}{2\pi}\int_{0}^{\lambda}z_{\{\widehat{w}^{\ast
}-\widehat{w}\}}^{T}(\omega)d\omega
\]
where $z_{\{\widehat{w}^{\ast}-\widehat{w}\}}^{T}(\omega)=\sum_{t=1}%
^{T}\{\widehat{w}^{\ast}(t)-\widehat{w}(t)\}e^{-\imath\omega t}$. Then we
have
\begin{align*}
\widehat{w}^{\ast}(t)-\widehat{w}(t) &  =\frac{1}{2\pi}\int_{0}^{2\pi
}e^{\imath\lambda t}dZ_{\{\widehat{w}^{\ast}-\widehat{w}\}}^{T}(\lambda)\\
&  =\frac{1}{T}\sum_{s=0}^{T-1}e^{\imath\frac{2\pi s}{T}t}z_{\{\widehat{w}%
^{\ast}-\widehat{w}\}}^{T}(2\pi s/T)\,,\quad t=1,\ldots,T\,,
\end{align*}
and direct substitution into equation \eqref{wdif} yields the equivalent
representations
\begin{align*}
\widehat{y}^{\ast}(t)-y(t) &  =\frac{1}{2\pi}\int_{0}^{2\pi}\sum_{j=0}%
^{t-1}\alpha_{j}^{(-\widehat{d})}e^{\imath\lambda(t-j)}dZ_{\{\widehat{w}%
^{\ast}-\widehat{w}\}}^{T}(\lambda)\\
&  =\frac{1}{T}\sum_{s=0}^{T-1}\sum_{j=0}^{t-1}\alpha_{j}^{(-\widehat{d}%
)}e^{\imath\frac{2\pi s}{T}(t-j)}z_{\{\widehat{w}^{\ast}-\widehat{w}\}}%
^{T}(2\pi s/T)\,,\quad t=1,\ldots,T\,.
\end{align*}

The Cauchy-Schwartz inequality now gives us
\begin{align*}
|\widehat{y}^{\ast}(t)-y(t)|  &  =\left\vert \frac{1}{T}\sum_{s=0}%
^{T-1}\left(  \sum_{j=0}^{t-1}\alpha_{j}^{(-\widehat{d})}e^{-\imath\frac{2\pi
s}{T} j}\right)  e^{\imath\frac{2\pi s}{T} t}z^{T}_{\{\widehat{w}^{\ast
}-\widehat{w}\}}(2\pi s/T)\right\vert \\
&  \leq\left(  \frac{1}{T}\sum_{s=0}^{T-1}\left\vert \sum_{j=0}^{t-1}%
\alpha_{j}^{(-\widehat{d})}e^{-\imath\frac{2\pi s}{T} j}\right\vert ^{2}
\right)  ^{1/2}\cdot\left(  \frac{1}{T}\sum_{s=0}^{T-1}|z^{T}_{\{\widehat{w}%
^{\ast}-\widehat{w}\}}(2\pi s/T )|^{2}\right)  ^{1/2}%
\end{align*}
and from Parseval's equality we have
\begin{align*}
\frac{1}{T}\sum_{s=0}^{T-1}\left\vert \sum_{j=0}^{t-1}\alpha_{j}%
^{(-\widehat{d})}e^{-\imath\frac{2\pi s}{T} j}\right\vert ^{2} = \sum
_{j=0}^{t-1}|\alpha_{j}^{(-\widehat{d})}|^{2}  &  \leq\sum_{j=0}^{\infty
}|\alpha_{j}^{(-\widehat{d})}|^{2}\\
&  = \frac{ \Gamma(1-2\widehat{d})}{(\Gamma(1-\widehat{d}))^{2}}<\frac{
\Gamma(\epsilon)}{(\Gamma(0.5(1+\epsilon) ))^{2}}\,.
\end{align*}
From the equality
\[
\frac{1}{T}\sum_{s=0}^{T-1}|z^{T}_{\{\widehat{w}^{\ast}-\widehat{w}\}}(2\pi
s/T )|^{2}=\frac{1}{T}\sum_{t=1}^{T}\{ \widehat{w}^{*}(t)-\widehat{w}(t)\}^{2}%
\]
it follows that
\[
\frac{1}{T}\sum_{t=1}^{T}|\widehat{y}^{\ast}(t)-y(t)|^{2}\leq\frac{
\Gamma(\epsilon)}{(\Gamma(0.5(1+\epsilon) ))^{2}}\frac{1}{T}\sum_{t=1}^{T}\{
\widehat{w}^{*}(t)-\widehat{w}(t)\}^{2}\,.
\]

The foregoing relationships hold for all possible pairs $(y(t),\widehat{y}%
^{\ast}(t))$, $t=1,\ldots,T$, with probability one with respect to
$(\Omega\otimes\Omega^{\ast},\mathfrak{F}\otimes\mathfrak{F}^{\ast},P(P^{\ast
}))$ and we can therefore conclude that
\[
E[E^{\ast}[\frac{1}{T}\sum_{t=1}^{T}|\widehat{y}^{\ast}(t)-y(t)|^{2}]]\leq
GE[E^{\ast}[\frac{1}{T}\sum_{t=1}^{T}\{\widehat{w}^{\ast}(t)-\widehat{w}%
(t)\}^{2}]],
\]
where $G=\Gamma(\epsilon)/(\Gamma(0.5(1+\epsilon)))^{2}$.\hfill\qed

\medskip

Now let $\widehat{\mathbf{s}}_{T}^{\ast}=(\widehat{s}_{1T}^{\ast}%
,\ldots,\widehat{s}_{mT}^{\ast})^{\prime}$, where $\widehat{s}_{iT}^{\ast
}=s_{i}(\widehat{y}^{\ast}(1),\ldots,\widehat{y}^{\ast}(T))$, $i=1,\ldots,m$,
denote the value of the statistic of interest when calculated from a
pre-filtered sieve bootstrap realization. Let $\widehat{\psi}_{T}^{\ast
}(\boldsymbol{\tau})=E^{\ast}[\exp(\imath\boldsymbol{\tau}^{\prime
}\widehat{\mathbf{s}}_{T}^{\ast})]$ and set $\widehat{\bzeta}_{T}^{\ast
}=\widehat{\mathbf{V}}_{T}^{\ast-1/2}T^{-\half}(\widehat{\mathbf{s}}_{T}%
^{\ast}-E^{\ast}[\widehat{\mathbf{s}}_{T}^{\ast}]) $ where
$\widehat{\mathbf{V}}_{T}^{\ast}=T^{-1}E^{\ast}\left[  (\widehat{\mathbf{s}%
}_{T}^{\ast}-E^{\ast}[\widehat{\mathbf{s}}_{T}^{\ast}])(\widehat{\mathbf{s}%
}_{T}^{\ast}-E^{\ast}[\widehat{\mathbf{s}}_{T}^{\ast}])^{\prime}\right]  $.

\begin{lemma}
\label{characff} Suppose that the process $y(t)$ satisfies Assumptions
\ref{Ass1} and \ref{Ass2}, and that the statistic $\mathbf{s}_{T}$ satisfies
Assumption \ref{linStat}. Then for all $\widehat{d}$ such that $\widehat{d}%
-d\in N_{\delta_{T}}$ where $\delta_{T}\rightarrow0$ as $T\rightarrow\infty$,
and $\widehat{d}\in(-0.5,0.5(1-\epsilon))$ where $\epsilon>0$, $E[E^{\ast
}[\Vert\widehat{\mathbf{s}}_{T}^{\ast}-\mathbf{s}_{T}\Vert^{2}]]=\exp
(2\delta_{T}\log T)o(T^{-1+\beta})$ for all $\beta>0$. Furthermore, if
$\delta_{T}\log T\rightarrow0$ as $T\rightarrow\infty$ then for all $\btau$
such that $\Vert\btau\Vert\leq T^{\beta/2}$, it follows that $|\widehat{\psi
}_{T}^{\ast}(\btau)-\psi_{T}(\btau)|=\exp(\delta_{T}\log T)o_{p}%
\{T^{-1/2+\beta}\}$ uniformly in $\btau$.
\end{lemma}

\noindent\textbf{Proof}: Since $\mathbf{s}_{T}$ satisfies Assumption
\ref{linStat} there exists a constant $M<\infty$ such that $\Vert
\widehat{\mathbf{s}}_{T}^{\ast}-\mathbf{s}_{T}\Vert^{2}\leq mMT^{-1}\sum
_{t=1}^{T}|(\widehat{y}^{\ast}(t)-y(t)|^{2}$, from which it immediately
follows that $E[E^{\ast}[\Vert\widehat{\mathbf{s}}_{T}^{\ast}-\mathbf{s}%
_{T}\Vert^{2}]]\leq mME[E^{\ast}[T^{-1}\sum_{t=1}^{T}|(\widehat{y}^{\ast
}(t)-y(t)|^{2}]]$. But by Proposition \ref{prefil} $E[E^{\ast}[T^{-1}%
\sum_{t=1}^{T}|(\widehat{y}^{\ast}(t)-y(t)|^{2}]]\leq GE[E^{\ast}[T^{-1}%
\sum_{t=1}^{T}|(\widehat{w}^{\ast}(t)-\widehat{w}(t)|^{2}]]$ where $G<\infty$,
and a repetition of the argument used in the proof of Lemma \ref{mbound} shows
that $E[E^{\ast}[|(\widehat{w}^{\ast}(t)-\widehat{w}(t)|^{2}%
]]=o(T^{-(1-2\widehat{\delta}^{\prime})+\beta})$ where $\widehat{\delta
}^{\prime}=\max\{0,d-\widehat{d}\}<\delta_{T}$ for all $\beta>0$. We are
therefore lead to the conclusion that $E[E^{\ast}[\Vert\widehat{\mathbf{s}%
}_{T}^{\ast}-\mathbf{s}_{T}\Vert^{2}]]=\exp(2\delta_{T}\log T)o(T^{-1+\beta}%
)$. This proves the first part of the lemma. The proof of the second part of
the lemma now follows that used in Lemma \ref{characf} in an obvious manner.
$\hfill\qed$

\begin{theorem}
\label{BTf} Suppose that the statistic $\mathbf{s}_{T}$ satisfies Assumption
\ref{linStat} and Assumption \ref{chfAss} with $r\geq4$ when calculated from
any process $y(t)$ that satisfies Assumptions \ref{Ass1} and \ref{Ass2}. Then
for all $\widehat{d}$ such that $\widehat{d}-d\in N_{\delta_{T}}$ where
$\delta_{T}\log T\rightarrow0$ as $T\rightarrow\infty$,
\[
\sup_{\mathbf{z}}|\textmd{P}^{*}(\widehat{\bzeta}^{*}_{T}\leq\mathbf{z}%
)-\textmd{P}(\bzeta_{T}\leq\mathbf{z})|=\exp(\delta_{T}\log T)O_{p}%
(T^{-1+\beta})
\]
for all $\beta>0$.
\end{theorem}

\noindent\textbf{Proof}: Apart from minor notational changes and an allowance
for the filtering that occurs at Steps PFSBS1 and PFSBS3, the argument leading
from Lemma \ref{characff} to Theorem \ref{BTf} is almost identical to that
leading from Lemma \ref{characf} to Theorem \ref{BT}. The details are
therefore omitted. $\hfill\qed$

\medskip

In practice, of course, the preliminary estimate $\widehat{d}$ will be
constructed from the data, and from Theorem \ref{BTf} we can see that if
$\widehat{d}-d\in N_{\delta_{T}}$ as $T\rightarrow\infty$, where $\delta
_{T}\log T\rightarrow0$, then the error of the pre--filtered sieve bootstrap
will be $O_{p}(T^{-1+\beta})$ for all $\beta>0$. Thus, if $|\widehat{d}-d|\log
T\rightarrow0$ $a.s.$ as $T\rightarrow\infty$ then the pre--filtered sieve
bootstrap will achieve a convergence rate arbitrarily close to the rate
obtained with simple random samples.

To establish the required convergence result for $\widehat{d}$ requires the
establishment of both consistency and the appropriate (limiting) tail
behaviour for the standardized estimator $N^{1/2}(\widehat{d}-d)$, where $N$
is a monotonically increasing function of $T$ such that $N/T\rightarrow0$ as
$T\rightarrow\infty$. In particular, if $N^{1/2}(\widehat{d}-d)$ were
$\mathbb{N}(0,\upsilon)$ then it would follow from the tail area properties of
the normal distribution that $\lim_{T\rightarrow\infty}P(|\widehat{d}%
-d|>\epsilon N^{-1/2+\delta})\leq\exp(-\epsilon^{2}N^{2\delta}/2\upsilon)$ for
any $\delta$, $0<\delta<0.5$ and $\epsilon>0$. Since $\exp(-\epsilon
N^{\delta}/2\upsilon)<|r|^{N^{\delta}}$ for all $r$ such that $\exp
(-\epsilon/2\upsilon)<|r|<1$ we could then conclude from the Borel-Cantelli
lemma that $N^{1/2-\delta}|\widehat{d}-d|$ converged to zero almost surely and
hence that $|\widehat{d}-d|\log T=o(1)$ $a.s.$ if $\log T/N^{1/2-\delta
}\rightarrow0$. Note that asymptotic Gaussianity (associated with a $\sqrt{N}%
$--CAN estimator of $d$) would not be sufficient here, as departures of
$N^{1/2}(\widehat{d}-d)$ from zero that are negligible in the sense of weak
convergence need not be so for large-deviation probabilities. Large-deviation
type results can, of course, be formally established on a case by case basis.
In particular, a corollary of \citet[][Lemma
5.8]{giraitis:robinson:2003} is that the semi-parametric local Whittle (SPWL)
estimator satisfies $P(|\widehat{d}-d|\log T>\epsilon)=o(N^{-p})$, where
$p>1/\epsilon$ and $N$, the bandwidth, satisfies $T^{\epsilon}<N<T^{1-\epsilon
}$ for some $\epsilon>0$.

In the simulation exercise that follows we apply a PFSBS algorithm based on a
pre-filtering value of $d$ that is produced by bias correcting the SPWL
estimator. The correction incorporates a combination of the analytical
adjustment of \cite{andrew:sun:2004} and a sieve bootstrap-based bias
adjustment, the latter justified on the basis of the Edgeworth expansion of
\citet{giraitis:robinson:2003}. In support of this choice of pre-filtering
value we invoke the Monte Carlo evidence in \cite{poskitt:martin:grose:2012}
that demonstrates the accuracy of (different versions of) a bias-adjusted SPWL
estimator, most notably in comparison with the raw SPWL estimator; see also
the discussion in \citet{nielsen:frederiksen:2005}. The simulation design
adopted in the current paper is identical to that adopted in
\citeauthor{poskitt:martin:grose:2012} and the bias-adjusted SPWL estimator
that minimized mean squared error across the Monte Carlo replications there,
in any given design setting, is used here as the pre-filter.

\section{Simulation Exercise\label{sim}}

In this section we examine the performance of the sieve bootstrap techniques
via a simulation experiment. Specifically, we investigate the accuracy with
which both the raw and pre-filtered sieve algorithms approximate the sampling
distributions of the sample mean, $\bar{y}_{T}$, and the $kth$-order sample
autocorrelation coefficient, $\hat{\rho}(k)$, for $k=1,$ $3,$ $6$ and $9$.

Regarding $\bar{y}_{T}=\sum_{t=1}^{T}y(t)/T$, various properties of this
statistic are well known, and in the investigation of any bootstrap procedure
an examination of its ability to mimic these is a natural focal point. In
particular, theoretical (asymptotic) properties notwithstanding, it is of
interest to investigate the nature of the finite sampling performance of the
sieve-based estimators of this important sampling distribution, and to
document the extent of the improvement yielded by the pre-filtering process.
The characteristics that are of particular interest in the context of
fractionally integrated data are, first, that
\begin{equation}
Var[\bar{y}_{T}]=\frac{1}{T}\sum_{k=1-T}^{T-1}(1-\frac{|k|}{T})\gamma
(k),\label{varybar}%
\end{equation}
second, that $Var[\bar{y}_{T}]\sim T^{2d-1}\omega^{2}$ where
\begin{equation}
\omega^{2}=\frac{\{\sigma\kappa(1)\}^{2}\Gamma(1-2d)}{(1+2d)\Gamma
(1+d)\Gamma(1-d)}\label{omega}%
\end{equation}
as $T\rightarrow\infty$, and third, that the re-normalized mean $T^{1/2-d}%
(\bar{y}_{T}-\mu)\overset{\mathcal{D}}{\rightarrow}\mathbb{N}(0,\omega^{2})$,
where $E[y_{t}]=\mu;$ see \citet[][Theorem 8]{hosking:1996}. In the case where
the simulated data is Gaussian all semi-invariants of $\bar{y}_{T}$ of order
greater than two are zero, of course, and the terms in the Edgeworth expansion
in \eqref{Edge1} beyond the first are null. Given knowledge of the true
sampling variance of the mean in \eqref{varybar}, the representativeness of
the Monte Carlo (MC) distribution, the relevance of the asymptotic
approximation and the accuracy of the bootstrap methods can all be assessed
against the exact Gaussian sampling
distribution.\footnote{\cite{andrews:lieberman:marmer:2006} remark that in the
Gaussian case \textquotedblleft the sample mean is an unbiased estimator of
$\mu$ with an exact normal distribution, which can be used to develop
inference concerning $\mu$\textquotedblright, but they make no mention of
issues associated with estimating the sampling variance of $\bar{y}_{T}$. We
should perhaps point out that the sample mean is not the best linear unbiased
estimator of $\mu$ for a fractional process, see \citet{adenstedt:1974}.
\citet[][Theorem
5.2]{adenstedt:1974} presents an alternative estimator that is asymptotically
efficient, albeit infeasible in practice because it is a function of the
unknown $d$. We thank the Editor for bringing this paper to our attention.}

With regard to the $k^{th}$ sample autocorrelation coefficient, defined here
as\footnote{There are several closely-related ways to define the sample
autocorrelation; we have used Hosking's (1996) specification in our work
here.}
\begin{equation}
\widehat{\rho}(k)=\frac{%
{\textstyle\sum\nolimits_{t=1}^{T-k}}
(y(t)-\bar{y}_{T})(y(t+k)-\bar{y}_{T})}{%
{\textstyle\sum\nolimits_{t=1}^{T}}
(y(t)-\bar{y}_{T})^{2}}, \label{phat}%
\end{equation}
the finite sample distribution under long memory is unknown. However, the
relevant asymptotic results are documented in \cite{hosking:1996}, with
asymptotic normality shown to hold for the appropriately standardized
statistic for $d\leq0.25,$ and Hosking's \textquotedblleft
modified\textquotedblright\ Rosenblatt distribution being the relevant
limiting distribution for $0.25<d<0.5.$ Hence, it is of interest to explore
the performance of the sieve-based techniques in replicating (Monte Carlo
estimates of) the finite sampling distributions in the two regions of the
parameter space in which the asymptotic behaviour of $\widehat{\rho}(k)$ differs.

Further, \cite{lieberman:rousseau:zucker:2001} \textrm{\ (\textquotedblleft
LRZ\textquotedblright\ hereafter)} provide the analytical details of the
Edgeworth expansion for the distribution of $\sqrt{T}\left(  \widehat{\rho
}_{0}(k)-\rho(k)\right)  $, where
\begin{equation}
\widehat{\rho}_{0}(k)=\frac{%
{\textstyle\sum\nolimits_{t=1}^{T-k}}
y(t)y(t+k)}{%
{\textstyle\sum\nolimits_{t=1}^{T}}
y(t)^{2}} \label{phat0}%
\end{equation}
and the mean $\mu$ is known to be zero. As well as being derived for a
statistic that both assumes and imposes a known zero mean, the LRZ expansion
is dependent on the true (unknown) values of the ARFIMA parameters, as well as
being valid only for very small values of $d$; all such features limiting its
empirical applicability. However, in the current experimental setting it
serves as a very useful check of the performance of the bootstrap method. A
good match in the case where the expansion is applicable suggests that the
bootstrap method(s) may also perform well in the parts of the parameter space
in which the Edgeworth approach is inapplicable, and in empirical scenarios in
which parameters are (of course) unknown. For convenience, and to facilitate
the reproducibility of our results, we document in the Appendix the details of
the Edgeworth formula applied here.

\subsection{Simulation Design\label{design}}

Data are simulated from a zero-mean Gaussian ARFIMA$(p,d,q)$ process, with
autoregressive lag order $p=1$ and moving average lag order $q=0,$
\begin{equation}
(1-L)^{d}(1-\phi L)y(t)=\varepsilon(t)\,, \label{arfima}%
\end{equation}
where $1-\phi z$ is the operator for a stationary AR(1) component and
$\varepsilon(t)$ is zero-mean Gaussian white noise. The theoretical
autocovariance function (ACF) for this process can be computed using the
procedures of \cite{sowell:1992}. The process in (\ref{arfima}) is simulated
$R=1000$ times for $d=0.0,0.2,0.3,0.4$; $\phi=0.3$ and $0.6;$ and for various
sample sizes $T$, via the Levinson recursion applied to the ACF of the desired
ARFIMA$(1,d,0)$ process and the generated pseudo-random $\varepsilon(t)$
\citep[see, for instance,][\S5.2]{brockwell:davis:1991}. The ARFIMA ACF for
given $T$, $\phi$ and $d$ is calculated using Sowell's
\citeyearpar{sowell:1992} algorithm as modified by \cite{doornik:ooms:2003}.

For each realization $r$ of the process we compute the relevant statistic,
$s_{T,r}$, plus $B=1000$ estimates $s_{T,r(b)}^{\ast}$, constructed using
$b=1,\ldots,B$ bootstrap re-samples obtained via the relevant bootstrap
algorithm. Each realized value $s_{T,r}$ thus has associated with it a
\textquotedblleft bootstrap distribution\textquotedblright\ based on the $B$
bootstrap resamples $s_{T,r(b)}^{\ast}$, $b=1,\ldots,B,$ with each such
distribution serving as an estimate of the sampling distribution of $s_{T}$.
In order to assess the $R$ bootstrap distributions against a comparator
distribution of $s_{T}$ -- whether that be the known finite sample
distribution (as in the case of $\overline{y}_{T}$), the finite sample
distribution estimated from the Monte Carlo draws, or an Edgeworth
approximation -- we first compute an \textquotedblleft
average\textquotedblright\ bootstrap distribution by sorting the $B$ bootstrap
draws for each MC replication into ascending order, then averaging these
ordered bootstrap values across the Monte Carlo draws. The $B$ averaged draws
are then used to produce a kernel density estimate, which we refer to as the
average\ bootstrap distribution. This bootstrap estimate, when plotted against
(a representation of) the true distribution of $s_{T}$, allows for a direct
visual assessment of the overall accuracy of the bootstrap
distributions\footnote{To document the extent of the variation in the
bootstrap samples (and hence density estimates) \textit{across} Monte Carlo
draws, we also produced kernel density estimates based on the quartiles of the
ordered bootstrap iterates.\textbf{\ }That is, if $s_{T,r(b_{j})}^{\ast},$
$r=1,\ldots,R$, is the set of $R$ $j^{th}$-largest bootstrap values, for which
we calculate the $q^{th}$ quantile, then the collection of $B$ such quantiles
is an estimate of the \textquotedblleft distribution\textquotedblright\ to the
left of which proportion $q$ of the $R$ BS distributions lie. Inspection of
these bootstrap quantiles did not indicate a great deal of variation in the
bootstrap distributions across Monte Carlo draws.}.

Following common practice \citep[][\S 3]{politis:2003}, the order of the
autoregressive approximation in the sieve was set to $h=\hat{h}_{T}%
=\mathrm{argmin}_{h=0,1,\ldots,M_{T}}\log(\hat{\sigma}_{h}^{2})+2h/T$, where
$\hat{\sigma}_{h}^{2}$ denotes the residual mean square obtained from an
$AR(h)$ model and $M_{T}=[(\log T)^{2}]$. Let $\bar{h}_{T}=\mathrm{argmin}%
_{h=0,1,\ldots,M_{T}}L_{T}(h)$ where $L_{T}(h)=(\sigma_{h}^{2}-\sigma
^{2})+h\sigma^{2}/T$ and $\sigma^{2}$ and $\sigma_{h}^{2}$ are as defined in
(\ref{Ass1}) and (\ref{Varh}) respectively. The function $L_{T}(h)$ was
introduced by \citet{shibata:1980} as a figure of merit and the $AR(\hat
{h}_{T})$ model is asymptotically efficient in the sense that $L_{T}(\hat
{h}_{T})=L_{T}(\bar{h}_{T})\{1+o(1)\}$ $a.s.$ as $T\rightarrow\infty$
\citep[][Theorem 9]{poskitt:2007}. It follows that $\hat{h}_{T}/\bar{h}%
_{T}\rightarrow1$ $a.s.$ as $T\rightarrow\infty$, so as $T$ increases $\hat
{h}_{T}$ behaves almost surely like a deterministic sequence that satisfies
the previous technical requirements. We can therefore conclude that, although
the use of $AIC$ introduces an added element of randomness, the results of
Section 3 and Section 4 -- in which $h$ is treated as fixed -- will still hold true.

In the case of $\bar{y}_{T}$ we supplement the graphical results by tabulating
the ratio of the average bootstrap estimate of the sampling variance of
$\bar{y}_{T}$ (for both bootstrap methods) to its true sampling variance as
per (\ref{varybar}). For the sample autocorrelations -- for which the finite
sample distribution is completely unknown -- we focus on the accuracy with
which the bootstrap methods reproduce the distribution as a whole; measuring
the \textquotedblleft closeness\textquotedblright\ of the averaged bootstrap
distribution (in any particular case) to the chosen comparator distribution
using three goodness of fit measures. For example, denoting the ordinates of
the Monte Carlo comparator and the (averaged) bootstrap-based probability
density functions (pdfs) at the $j^{th}$ (sorted) Monte Carlo realization
value as $p_{mc}(s_{j})$ and $p_{bs}(s_{j})$ respectively, we calculate
\[
RMSD=\sqrt{\frac{1}{R}\sum_{j=1}^{R}\left(  p_{mc}(s_{j})-p_{bs}%
(s_{j})\right)  ^{2}}%
\]
and the Kullback-Leibler divergence\textrm{\ }%
\[
KLD=\sum_{j=1}^{R}p_{mc}(s_{j})\ln\frac{p_{mc}(s_{j})}{p_{bs}(s_{j})}%
\simeq\frac{1}{R}\sum_{j=1}^{R}\ln\frac{p_{mc}(s_{j})}{p_{bs}(s_{j})},
\]
where this last follows given $s_{j}$ is by definition a random draw from the
Monte Carlo distribution of $s_{T}$. We also produce a \textquotedblleft
GINI\textquotedblright-style statistic by estimating the cumulative
distribution functions, $P_{mc}(s_{j})$ and $P_{bs}(s_{j}),$ $j=1,2,...,J,$ in
the same manner as the pdfs above, then calculating the area between the
resulting PP-plot and the line of equality via a numerical (trapezoidal)
estimate of%
\[
A=\int_{0}^{1}\left\vert P_{bs}(s)-P_{mc}(s)\right\vert dP_{mc}(s).
\]
The GINI coefficient is then just $2A.$

\subsection{Simulation results: sample mean}

\begin{figure}[h]
\centering\includegraphics[trim=10mm 70mm 10mm
80mm,width=6in]{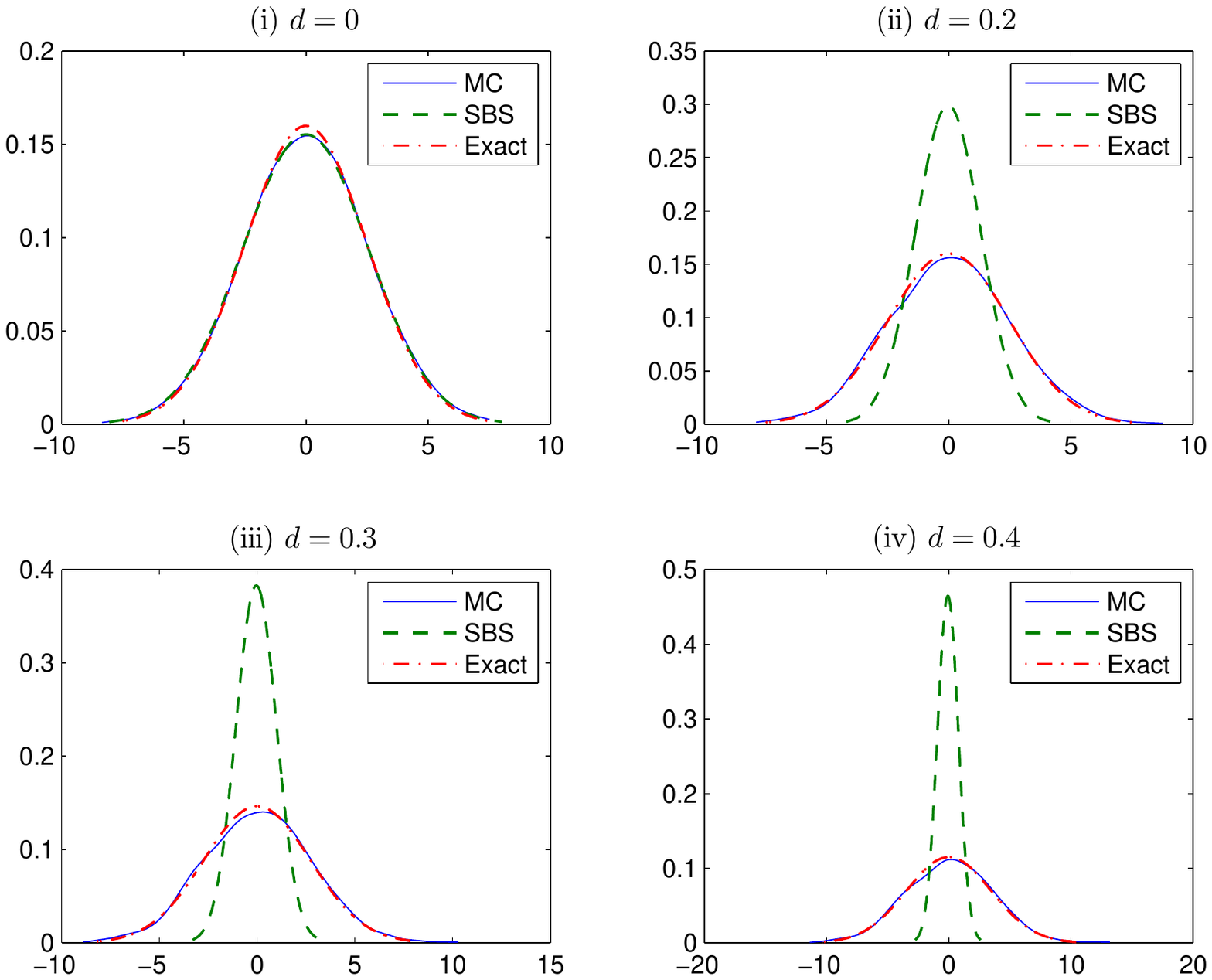}\caption{Densities of the re-normalized
sample mean under ARFIMA$(1,d,0)$ with $T=500$, $\phi=0.6$, and
$d=0,0.2,0.3,0.4$: Monte Carlo (MC); averaged (raw) sieve bootstrap (SBS) and
exact ($N(0,\overline{\omega}^{2}$)), where $\overline{\omega}^{2}%
=T^{1-2d}Var[\bar{y}_{T}]$, with $Var[\bar{y}_{T}]$ as given in (\ref{varybar}%
).}%
\label{RSBMean}%
\end{figure}

Figure \ref{RSBMean} graphs the distribution of $T^{1/2-d}(\bar{y}_{T}-\mu)$
(where $\mu=0$) observed across the Monte Carlo draws (denoted by MC), the
averaged (raw) sieve bootstrap distribution of $T^{1/2-d}(\bar{y}_{T}^{\ast
}-\bar{y}_{T})$ (denoted by SBS), and the exact $\mathbb{N}(0,\overline
{\omega}^{2})$ distribution, with $\overline{\omega}^{2}=T^{1-2d}Var[\bar
{y}_{T}],$ for $T=500$, $\phi=0.6$, and $d=0,0.2,0.3,0.4$. We have suppressed
the plot of the asymptotic $\mathbb{N}(0,\omega^{2})$ distribution of
$T^{1/2-d}(\bar{y}_{T}-\mu)$ since at this sample size it is virtually
indistinguishable from the exact. 

When $d=0$ we can see that all three distributions are very nearly identical.
When $d>0$, however, it is clear that the variance of $\bar{y}_{T} $ is
substantially underestimated by the bootstrap procedure. This result is
further confirmed by inspection of Table \ref{stdev:ratio}, which reports the
ratio of the average SBS estimate of the standard deviation of $\bar{y}_{T}$
to the exact standard deviation given by the square root of \eqref{varybar},
for $T=100$ and $500,$ and for $\phi=0.3$ and $0.6.$ The underestimation for
$d>0$ is very marked for both values of $\phi$ and both sample sizes with,
indeed, the degree of underestimation increasing with $\phi$ and there being
no uniform tendency for improvement as the sample size increases\footnote{The
mean and skewness of the re-nomalised difference $T^{1/2-d}(\bar{y}_{T}^{\ast
}-\bar{y}_{T})$ are close to zero, and the kurtosis is approximately 3. Thus
it is only the underestimation of variance that presents a problem. A similar
phenomenon with the block bootstrap was observed previously by
\cite{hesterberg:1997}. Hesterberg offers no explanation for its occurrence,
but simply suggests that estimating the variance of the sample mean is
substantially more difficult in the long memory case than it is for short
memory processes.}.%

\begin{table}[tbp] \centering
\caption{Standard deviation of $\bar{y}_{T}$: averaged SBS estimate as a
percentage of $\protect\sqrt{Var[\bar{y}_{T}]}$, with $Var[\bar{y}_{T}]$ as
defined in (\protect\ref{varybar}).} \label{stdev:ratio}%

\begin{tabular}
[c]{ccccccc}
&  &  & \multicolumn{4}{c}{}\\
&  &  & \multicolumn{4}{c}{$d$}\\\cline{4-7}
&  &  & 0.0 & 0.2 & 0.3 & 0.4\\
$\phi$ & $T$ &  &  &  &  & \\\cline{1-2}%
0.3 & 100 &  & 95.6\% & 57.2\% & 42.6\% & 28\%\\
& 500 &  & 99.2\% & 48.6\% & 35.1\% & 22.8\%\\
&  &  &  &  &  & \\
0.6 & 100 &  & 93.3\% & 60.2\% & 46\% & 31\%\\
& 500 &  & 99.1\% & 51.5\% & 36.8\% & 23.8\%\\
&  &  &  &  &  & \\
&  &  &  &  &  &
\end{tabular}
%

\end{table}%

The reason for the underestimation stems from the fact that the raw sieve
bootstrap variance is
\[
Var^{\ast}[\bar{y}_{T}^{\ast}]=\frac{1}{T}\sum_{k=1-T}^{T-1}(1-\frac{|k|}%
{T})\bar{\gamma}_{h}(k),
\]
where $\bar{\gamma}_{h}(k)=\widehat{\gamma}(k)$, $k=0,1,\ldots,h$, and
$\sum_{j=0}^{h}\bar{\phi}_{h}(j)\bar{\gamma}_{h}(k-j)=0$, $k=h+1,\ldots$, with
$\widehat{\gamma}(k)=\frac{1}{T-k}%
{\textstyle\sum\nolimits_{t=1}^{T-k}}
(y(t)-\bar{y}_{T})(y(t+k)-\bar{y}_{T})$; and \cite{hosking:1996} shows that
the $\widehat{\gamma}(k)$ can have substantial negative bias relative to the
corresponding true values even for moderate to large samples, particularly
when $d$ is large.

\begin{figure}[h]
\centering
\includegraphics[trim=10mm 100mm 10mm 90mm,width=6in]{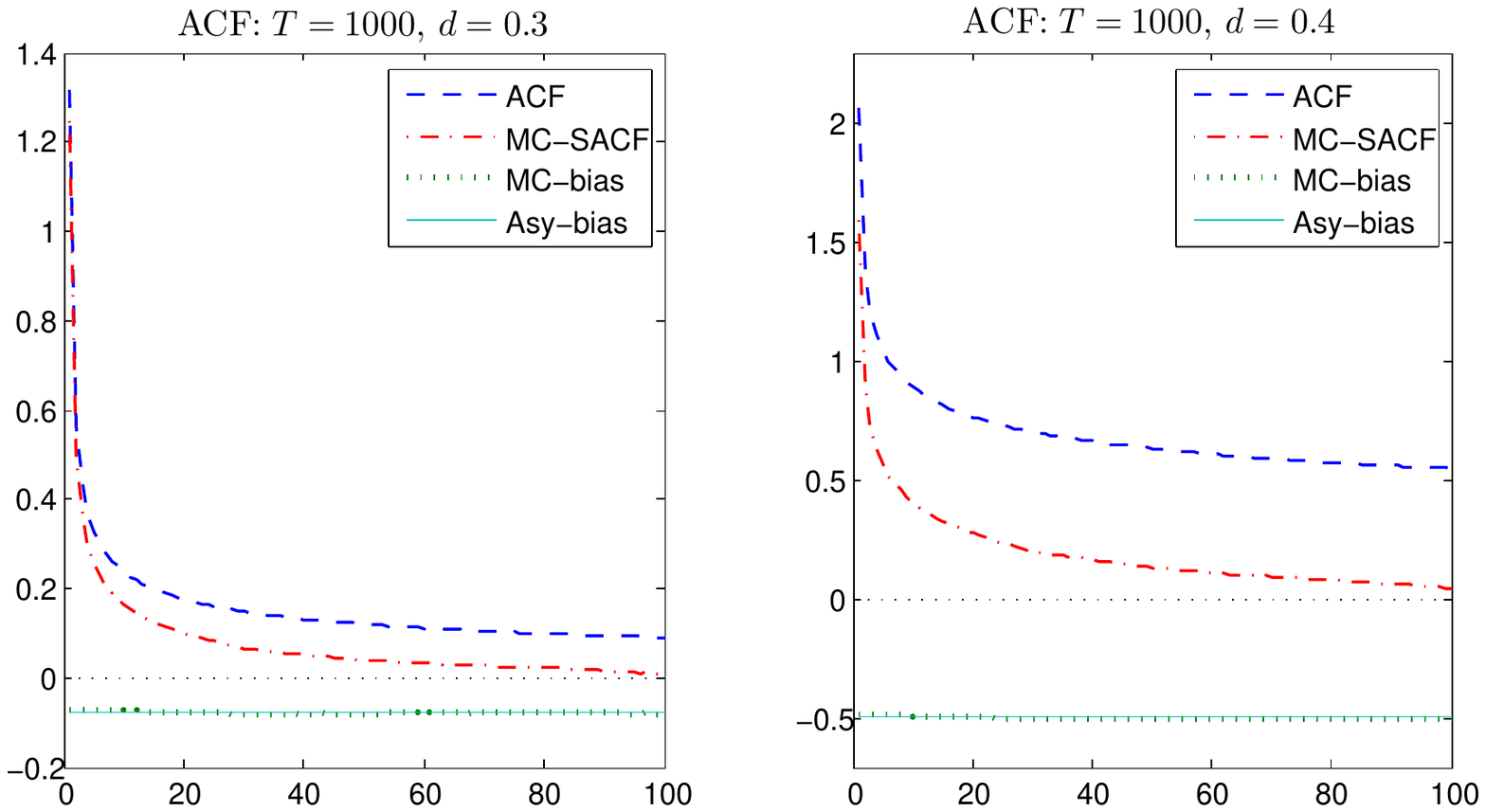}\caption{Theoretical
autocovariance function (ACF); Monte-Carlo estimate of the expected sample
autocovariance function (MC-SACF); Monte-Carlo estimate of the bias in the
SACF (MC-Bias); and the asymptotic bias as per (\ref{autobias}) (Asy-Bias),
under fractional noise with $d=0.3,0.4$ and $T=1000$. }%
\label{ACFfig}%
\end{figure}

This phenomenon is illustrated in Figure \ref{ACFfig}, which depicts the
theoretical autocovariance function and the value of $\widehat{\gamma}(k)$ for
$k=0,\ldots,100$ obtained from samples of size $T=1000$, computed from two
fractional noise (ARFIMA$(0,d,0)$) processes with $d=0.3$ and $d=0.4$, and
averaged across the $R$ replications. \citet[][Theorem 3]{hosking:1996}
provides the following formula for the asymptotic bias of the autocovariances
\begin{equation}
E[\widehat{\gamma}(k)-\gamma(k)]\sim-\omega^{2}T^{2d-1}\,, \label{autobias}%
\end{equation}
(with $\omega^{2}$ as defined in \ref{omega}), which depends on $d$ but is
independent of $k$. This feature is reflected in the simulated sample bias,
computed as the difference between the mean of the simulated sample
autocovariances at each lag $k$ and the corresponding true $\gamma(k)$. For
$T=1000$ this estimated bias is in close accord with the asymptotic
approximation in \eqref{autobias}, as can be seen in Figure \ref{ACFfig}%
\footnote{Noting that $\frac{1}{T}\sum_{k=1-T}^{T-1}(1-\frac{|k|}{T})\equiv1$,
it is apparent from equation \eqref{autobias} that the addition of $\omega
^{2}T^{2d-1}$ to the bootstrap variance would provide an asymptotically valid
(albeit empirically infeasible) correction that would compensate for the bias
of the $\widehat{\gamma}(k)$ and the underestimation of the true persistence
in the observed process.}.

It is of interest then to ascertain whether the feasible PFSBS algorithm, in
\textit{implicitly} producing more accurate estimates of the $\gamma(k)$ in
the process of yielding bootstrap draws of $\bar{y}_{T}$, (via the application
of the sieve to a shorter memory process) yields an estimated sampling
distribution for the mean with a variance that is closer to the theoretical
value. As noted in Section \ref{pfsbs}, the PFSBS algorithm is based on a
pre-filtering value of $d$ that is deemed to be \textquotedblleft
optimal\textquotedblright\ in the matching experimental design in
\cite{poskitt:martin:grose:2012}.

\begin{figure}[h]
\centering\includegraphics[trim=10mm 70mm 10mm
80mm,width=6in]{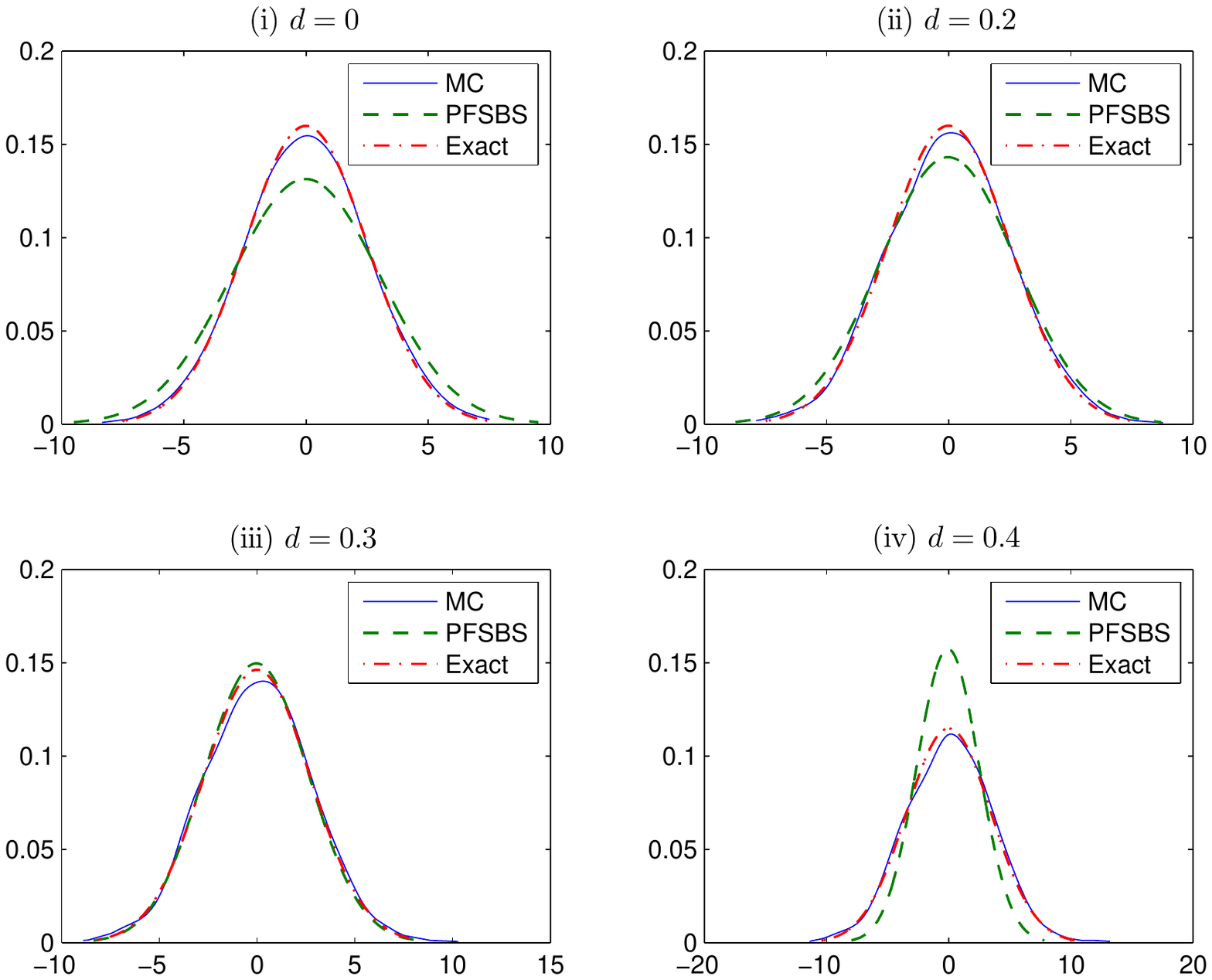}\caption{Densities of the re-normalized
sample mean under ARFIMA$(1,d,0)$ with $T=500$, $\phi=0.6$, and
$d=0,0.2,0.3,0.4$: Monte Carlo (MC); averaged pre-filtered sieve bootstrap
(PFSBS) and exact ($N(0,\overline{\omega}^{2}$)), where $\overline{\omega}%
^{2}=T^{1-2d}Var[\bar{y}_{T}]$, with $Var[\bar{y}_{T}]$ as given in
(\ref{varybar}).}%
\label{RSBMean2}%
\end{figure}

Figure \ref{RSBMean2} graphs the Monte Carlo distribution of $T^{1/2-d}%
(\bar{y}_{T}-\mu)$, the PFSBS distribution of $T^{1/2-d}(\bar{y}_{T}^{\ast
}-\bar{y}_{T})$, and the $\mathbb{N}(0,\overline{\omega}^{2})$ distribution,
for $T=500$, $\phi=0.6$, and $d=0,0.2,0.3,0.4$. We see that, despite a tendency to over-estimate $Var(\bar{y}%
_{T})$ for $d=0$ and under-estimate for large $d$ ($d=0.4$), the PFSBS results
are far superior to those associated with the raw SBS, and reasonably close
overall to the true sampling distribution. The averaged PFSBS estimate of
$\sqrt{Var\left(  \bar{y}_{T}\right)  }$ as a percentage of the true
$\sqrt{Var\left(  \bar{y}_{T}\right)  }$ is presented in Panel A of Table
\ref{PFBSstdev:ratio}, for the two values of $\phi$, $\phi=0.3$ and $0.6,$ and
for $T=100$ and $500.$ The reasonable accuracy observed visually in Figure
\ref{RSBMean2} for $\phi=0.6,$ for the larger sample size in particular, is
broadly replicated for $\phi=0.3$, $T=500$, augering well for the automated
use of the pre-filtering method in practice.%

\begin{table}[tbp] \centering
\caption{Standard deviation of $\bar{y}_{T}$: averaged PFSBS and FPSBS estimates as a
percentage of $\protect\sqrt{Var[\bar{y}_{T}]}$, with $Var[\bar{y}_{T}]$ as
defined in (\protect\ref{varybar}).} \label{PFBSstdev:ratio}%

\begin{tabular}
[c]{cccccc}
&  &  &  &  & \\
\multicolumn{6}{c}{Panel A: PFSBS\textbf{\ }}\\
&  &  &  &  & \\
&  & \multicolumn{4}{c}{$d$}\\\cline{3-6}
&  & $0.0$ & $0.2$ & $0.3$ & $0.4$\\
$\phi$ & $T$ &  &  &  & \\\cline{1-2}%
0.3 & 100 & 141.2\% & 125.3\% & 109.8\% & 84.3\%\\
& 500 & 116.6\% & 106.9\% & 93.4\% & 69.9\%\\
&  &  &  &  & \\
0.6 & 100 & 158.7\% & 142.9\% & 127.0\% & 100.4\%\\
& 500 & 117.1\% & 107.5\% & 94.0\% & 70.1\%\\
&  &  &  &  & \\
\multicolumn{6}{c}{Panel B: FPFBS\textbf{\ }}\\
&  &  &  &  & \\
&  & \multicolumn{4}{c}{$d$}\\\cline{3-6}
&  & $0.0$ & $0.2$ & $0.3$ & $0.4$\\
$\phi$ & $T$ &  &  &  & \\\cline{1-2}%
0.3 & 100 & 349.0\% & 191.1\% & 127.8\% & 74.3\%\\
& 500 & 573.4\% & 274.9\% & 171.6\% & 88.6\%\\
&  &  &  &  & \\
0.6 & 100 & 316.0\% & 169.5\% & 115.7\% & 69.7\%\\
& 500 & 582.9\% & 268.6\% & 162.5\% & 84.7\%\\
&  &  &  &  & \\
&  &  &  &  &
\end{tabular}
%

\end{table}%

As a final point here, it is of interest to ascertain the performance of the
PFSBS technique in which we simply \emph{assign} a value to $d$ with which to
pre-filter, rather than selecting a particular estimator for this
role.\footnote{The idea of imposing a \textquotedblleft
fixed\textquotedblright\ pre-filter arose out of a referee's comment on an
earlier version of the paper.} A fairly obvious choice is to set the
pre-filtering value ($d^{f}$ say) at $0.5;$ as the true $d$ (in the
experimental setting) is never greater than this it follows that imposing
$d^{f}=0.5$ results in a filtered series for which the effective fractional
integration is always negative, and the filtered process of intermediate
memory as a consequence. The estimates of the sieve parameters will therefore
converge at the best possible rate $O\left(  hT^{-1}\log T\right)  $ as per
Theorem \ref{consistentYW}, although $d^{f}$ will obviously \emph{not} satisfy
the convergence properties outlined in the discussion following Theorem
\ref{BTf}. We refer to this approach below as the \textquotedblleft fixed
pre-filtered bootstrap\textquotedblright\ (FPFBS).

\begin{figure}[h]
\centering\includegraphics[trim=10mm 70mm 10mm
80mm,width=6in]{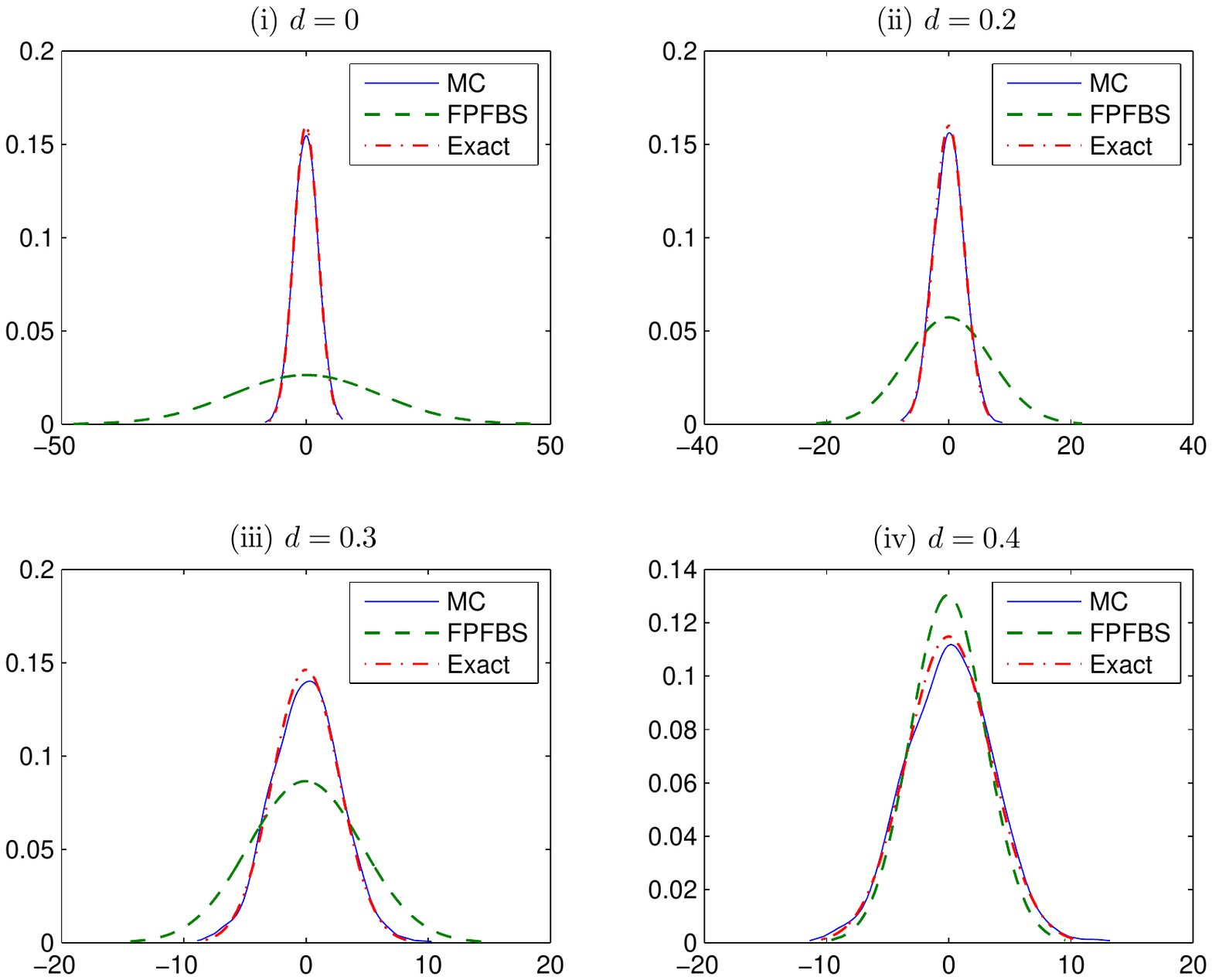}\caption{Densities of the re-normalized
sample mean under ARFIMA$(1,d,0)$ with $T=500$, $\phi=0.6$, and
$d=0,0.2,0.3,0.4$: Monte Carlo (MC); averaged fixed pre-filtered sieve
bootstrap (FPFBS) and exact ($N(0,\overline{\omega}^{2}$)), where
$\overline{\omega}^{2}=T^{1-2d}Var[\bar{y}_{T}]$, with $Var[\bar{y}_{T}]$ as
given in (\ref{varybar}).}%
\label{RSBMean3}%
\end{figure}

As we see from Figure \ref{RSBMean3}, the FPFBS, unsurprisingly, works
reasonably well when $d$ is large; i.e., for $d=0.4.$ For the smaller values
for $d$, on the other hand, it works very poorly, resulting in an averaged
bootstrap distribution for $\bar{y}_{T}$ that is a very inaccurate match for
the true distribution. In particular, as seen in panels (i) -- (iii) of the
Figure, and in Panel B of Table \ref{PFBSstdev:ratio}, the dispersion of the
FPFBS-based distribution is much larger than that of the exact distribution,
with the discrepancy increasing with the distance $\left\vert d-d^{f}%
\right\vert .$ In short, it appears that fixing the pre-filter is not useful
as a default setting, at least as regards estimating the distribution of
$\bar{y}_{T}$.

\subsection{Simulation results: sample autocorrelation\label{auto}}

\begin{figure}[h]
\centering
\includegraphics[trim=10mm 70mm 10mm 80mm,width=6in]{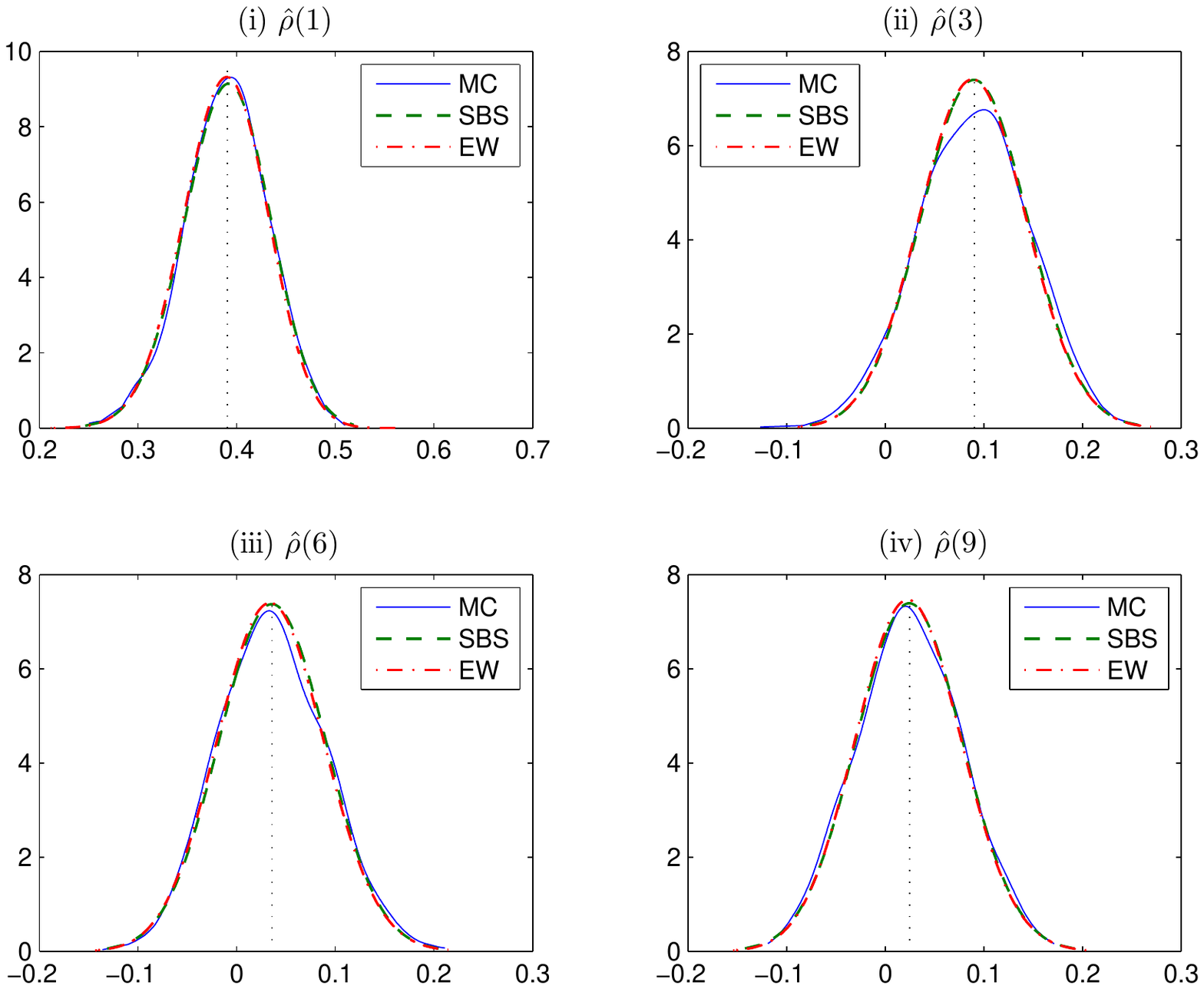}\caption{Densities
of the zero-mean sample autocorrelation coefficient $\hat{\rho}_{0}(k)$,
$k=1,3,6,9$: under ARFIMA$(1,d,0)$ with $T=500$, $d=0.08$, $\phi=0.3$: Monte
Carlo (MC); averaged (raw) sieve bootstrap (SBS); Edgeworth approximation
(EW). The vertical dotted line indicates the position of the true value of
$\rho(k)$, $k=1,3,6,9.$}%
\label{bsmcedge1}%
\end{figure}

\begin{figure}[h]
\centering
\includegraphics[trim=10mm 70mm 10mm 80mm,width=6in]{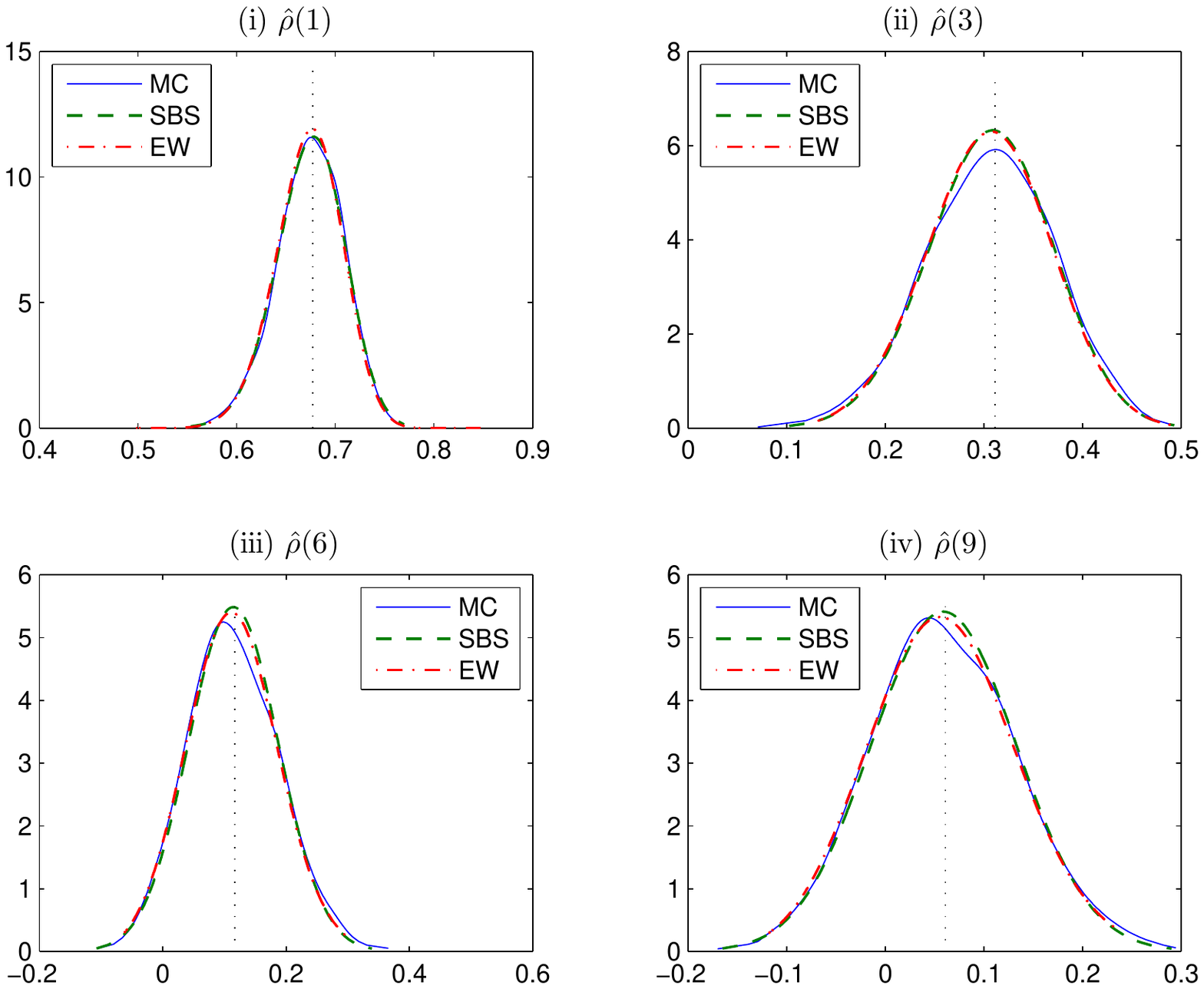}\caption{Densities
of the zero-mean sample autocorrelation coefficient $\hat{\rho}_{0}(k)$,
$k=1,3,6,9$: under ARFIMA$(1,d,0)$ with $T=500$, $d=0.08$, $\phi=0.6$: Monte
Carlo (MC); averaged (raw) sieve bootstrap (SBS); Edgeworth approximation
(EW). The vertical dotted line indicates the position of the true value of
$\rho(k)$, $k=1,3,6,9.$}%
\label{bsmcedge2}%
\end{figure}

We begin by plotting various estimates of the true finite sampling
distribution of $\widehat{\rho}_{0}(k),$ for $k=1,3,6$ and $9,$ where the
subscript $0$ is used to emphasize that a mean of zero (for $y_{t}$) is both
assumed and imposed in the calculation of the statistic (see \eqref{phat0}).
We consider this particular version of the autocorrelation coefficient(s) in
this initial exercise so as to enable the LRZ expansion (derived for this
version) to be used as a comparator. The expansion is valid for $d<0.1$ only
(see Appendix); hence we conduct the comparison for a value of $d$ in this
range: $d=0.08.$ Results for $\phi=0.3$ and $\phi=0.6$ are presented in
Figures \ref{bsmcedge1} and \ref{bsmcedge2} respectively, with $T=500$ in both cases.

As is evident from inspection of the two graphs, the (raw) SBS estimate of the
distribution of $\widehat{\rho}_{0}(k)$ is visually indistinguishable from the
Edgeworth distribution\footnote{The Edgeworth (EW) distribution plotted here
has been re-centered on the true $\rho(k)$, and rescaled to remove the
$\sqrt{T}$ normalization of the expansion. See the Appendix for details.},
with both being very similar to the Monte Carlo based estimate.\footnote{We
have reproduced results here based on 1000 replications in order to have all
results comparable throughout the paper. In particular, due to the
computational burden associated with the PFSBS methods, 1000 was a manageable
choice for a general replication number. However, the results documented in
Figures 5 and 6 have also been run using 10,000 replications, at which point
the Monte Carlo estimate of the pdf is visually indistinguishable from the
other two estimates.} As such, we conclude that when a finite sample
comparator is available (i.e. under the conditions required for that
comparator to be valid) the sieve bootstrap method is remarkably accurate.
This gives one confidence in the ability of the bootstrap to provide an
accurate result in the usual case in which such a comparator is unavailable.

\begin{figure}[h]
\centering
\includegraphics[trim=10mm 70mm 10mm 80mm,width=6in]{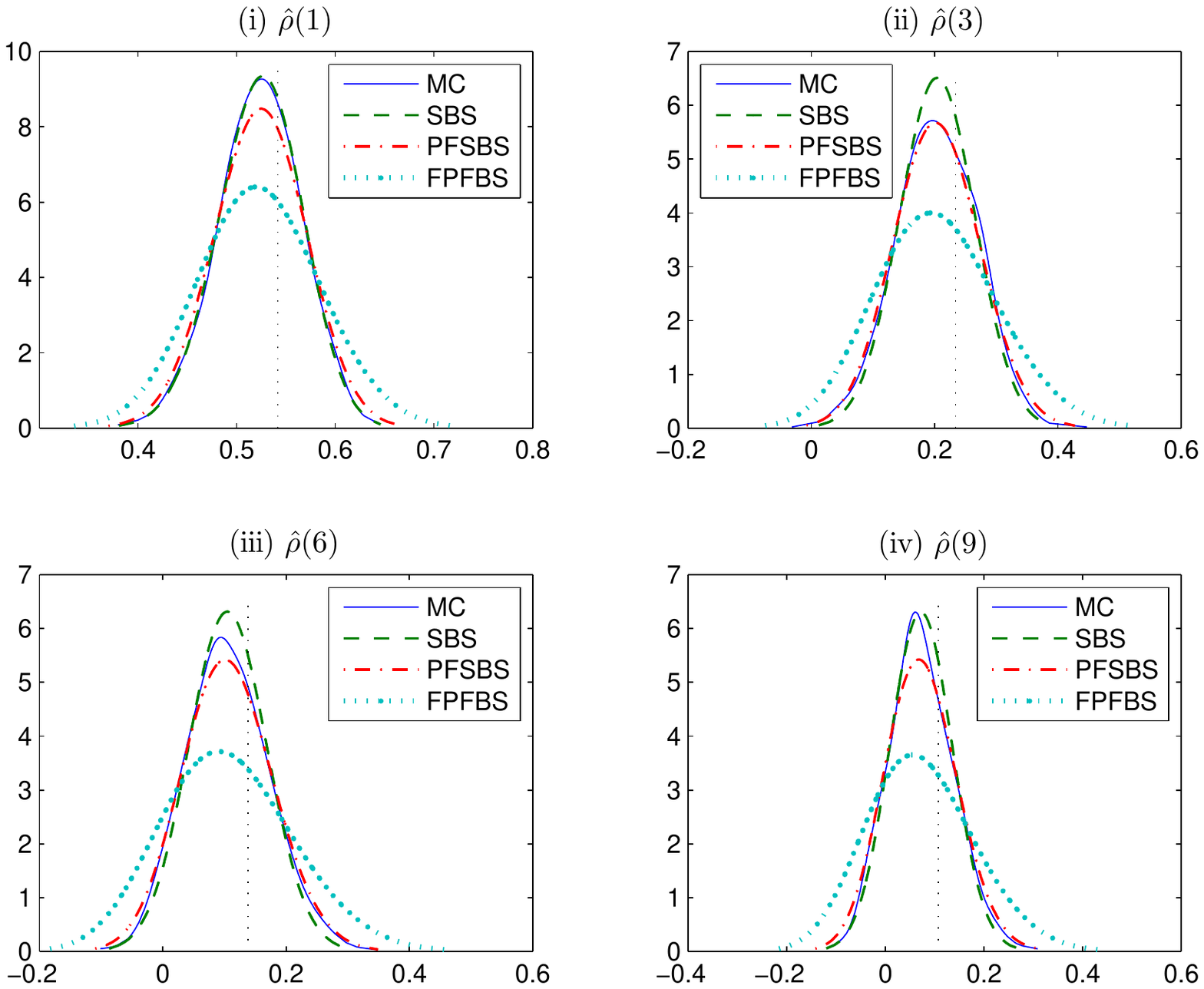}\caption{Densities
of the sample autocorrelation coefficient $\hat{\rho}(k)$, $k=1,3,6,9$: under
ARFIMA$(1,d,0)$ with $T=500$, $d=0.2$, $\phi=0.3$: Monte Carlo (MC); averaged
(raw) sieve bootstrap (SBS); averaged pre-filtered sieve bootstrap (PFSBS);
averaged fixed pre-filtered sieve bootstrap (FPFBS). The vertical dotted line
indicates the position of the true value of $\rho(k)$, $k=1,3,6,9.$}%
\label{bsmcfig1}%
\end{figure}

\begin{figure}[h]
\centering
\includegraphics[trim=10mm 70mm 10mm 80mm,width=6in]{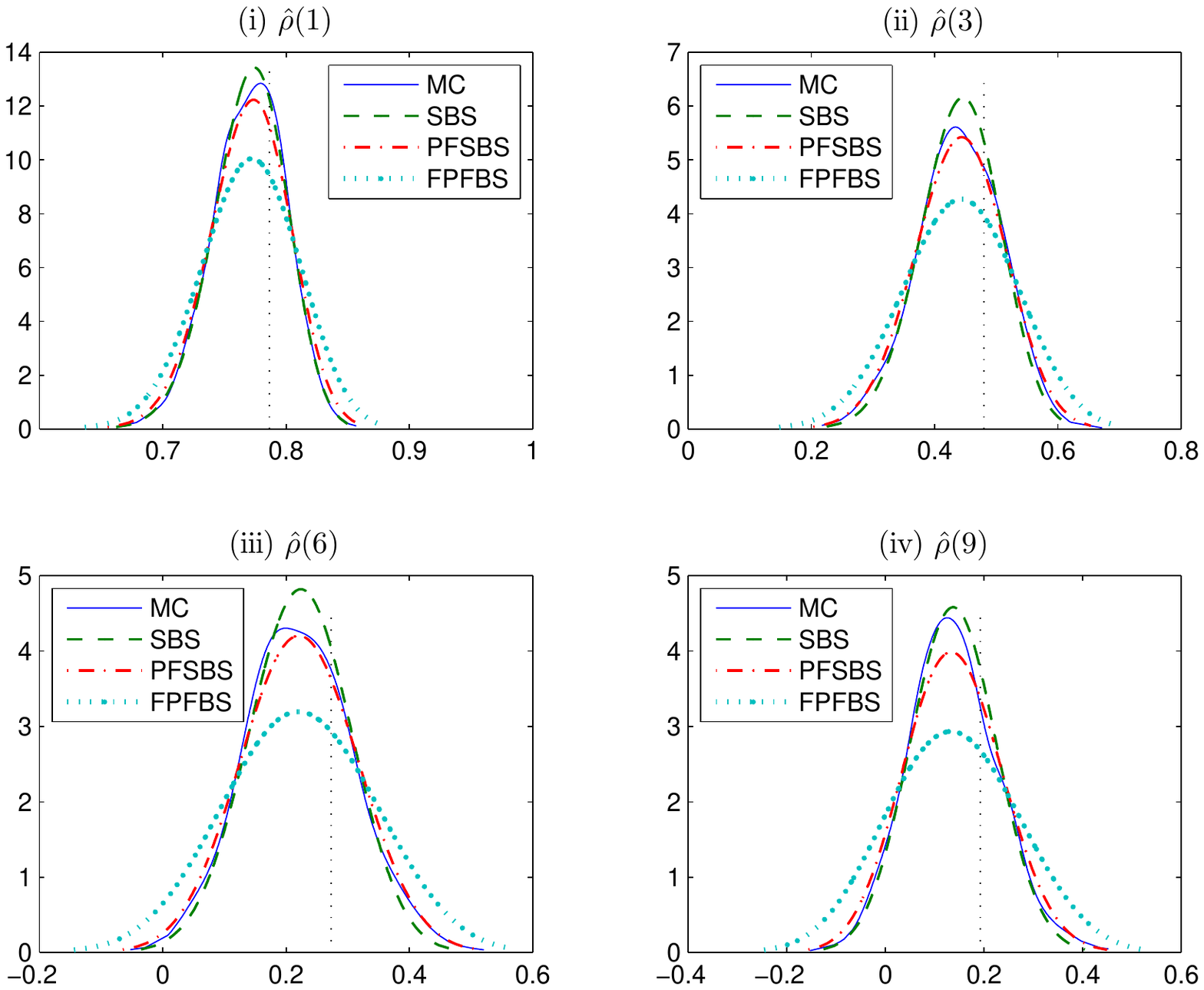}\caption{Densities
of the sample autocorrelation coefficient $\hat{\rho}(k)$, $k=1,3,6,9$: under
ARFIMA$(1,d,0)$ with $T=500$, $d=0.2$, $\phi=0.6$: Monte Carlo (MC); averaged
(raw) sieve bootstrap (SBS); averaged pre-filtered sieve bootstrap (PFSBS);
averaged fixed pre-filtered sieve bootstrap (FPFBS). The vertical dotted line
indicates the position of the true value of $\rho(k)$, $k=1,3,6,9.$}%
\label{bsmcfig2}%
\end{figure}

\begin{figure}[h]
\centering
\includegraphics[trim=10mm 70mm 10mm 80mm,width=6in]{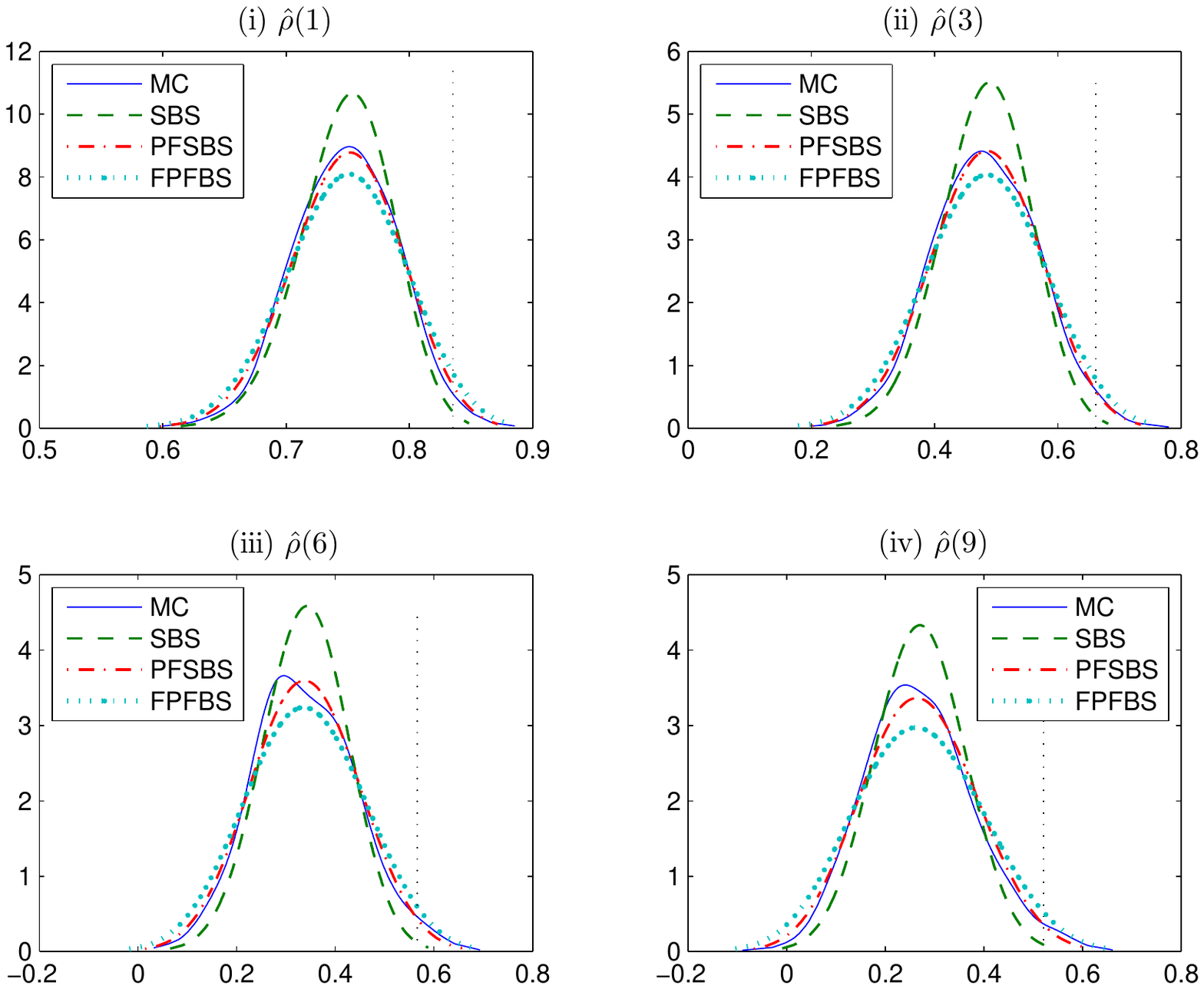}\caption{Densities
of the sample autocorrelation coefficient $\hat{\rho}(k)$, $k=1,3,6,9$: under
ARFIMA$(1,d,0)$ with $T=500$, $d=0.4$, $\phi=0.3$: Monte Carlo (MC); averaged
(raw) sieve bootstrap (SBS); averaged pre-filtered sieve bootstrap (PFSBS);
averaged fixed pre-filtered sieve bootstrap (FPFBS). The vertical dotted line
indicates the position of the true value of $\rho(k)$, $k=1,3,6,9.$}%
\label{bsmcfig3}%
\end{figure}

\begin{figure}[h]
\centering
\includegraphics[trim=10mm 70mm 10mm 80mm,width=6in]{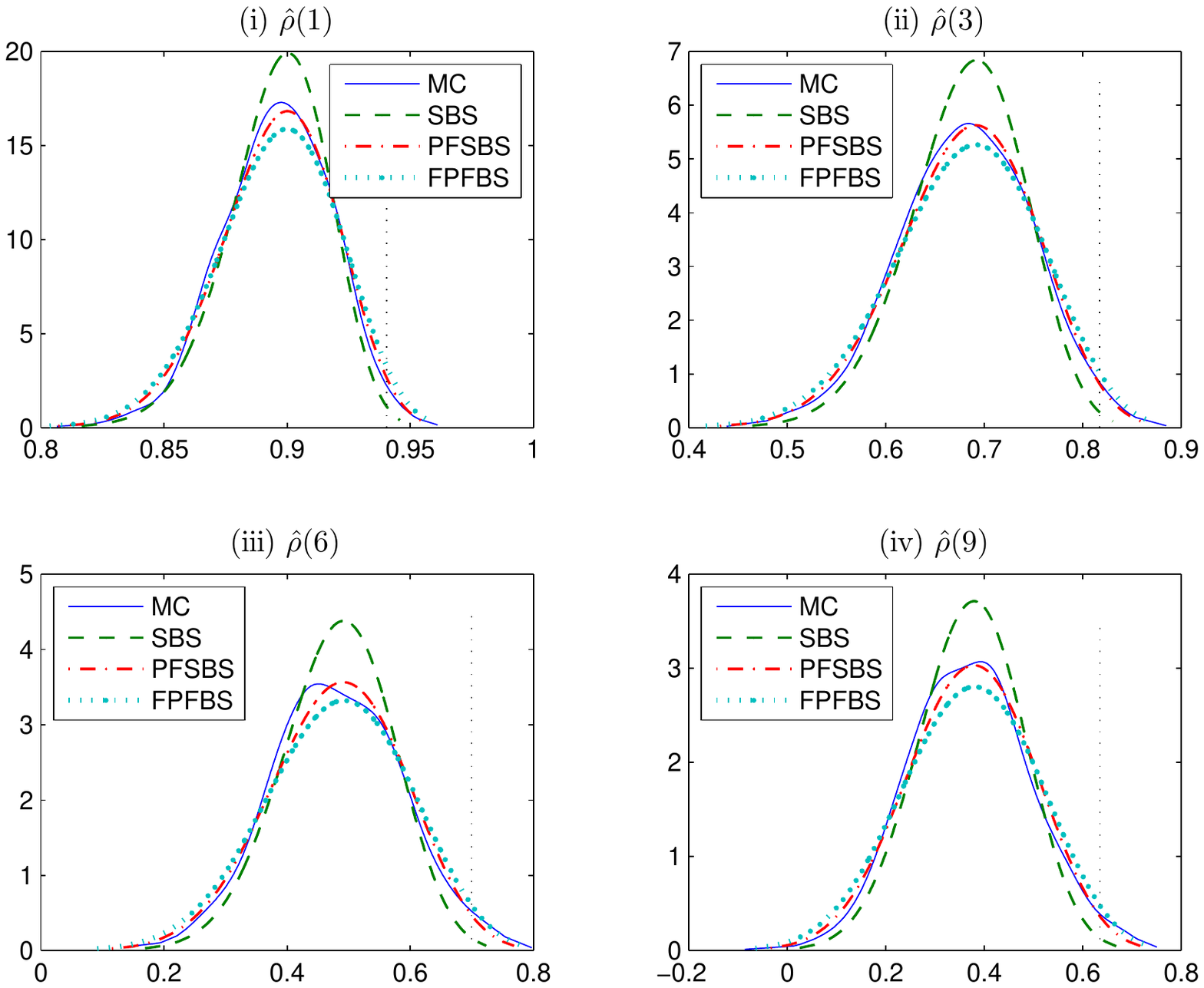}\caption{Densities
of the sample autocorrelation coefficient $\hat{\rho}(k)$, $k=1,3,6,9$: under
ARFIMA$(1,d,0)$ with $T=500$, $d=0.4$, $\phi=0.6$: Monte Carlo (MC); averaged
(raw) sieve bootstrap (SBS); averaged pre-filtered sieve bootstrap (PFSBS);
averaged fixed pre-filtered sieve bootstrap (FPFBS). The vertical dotted line
indicates the position of the true value of $\rho(k)$, $k=1,3,6,9.$}%
\label{bsmcfig4}%
\end{figure}

In Figures \ref{bsmcfig1}--\ref{bsmcfig4} we proceed to document the
performance of the two bootstrap methods, SBS and PFSBS, in regions of the
parameter space in which the Edgeworth expansion is not valid, and the only
comparator is the Monte Carlo-based estimate of the exact sampling
distribution. We also include plots of the average FPFBS distributions
calculated using, as in the previous section, $d^{f}=0.5.$ The distribution
for the sample autocorrelation coefficient in (\ref{phat}) is now the one
documented, for $k=1,$ $3,$ $6$ and $9$, and the two scenarios considered are
that in which asymptotic normality holds ($d\leq0.25),$ and that in which it
does not $(d>0.25)$, with the modified Rosenblatt distribution being the
relevant limiting form in the latter case. Specifically, in Figures
\ref{bsmcfig1} and \ref{bsmcfig2}, $d=0.2$ and $\phi=0.3$ and $\phi=0.6$
respectively, whilst in Figures \ref{bsmcfig3} and \ref{bsmcfig4}, $d=0.4$ and
$\phi=0.3$ and $\phi=0.6$ respectively. In order to supplement these graphical
results, the measures of fit (as described in Section \ref{design}) are
recorded in both panels of Table \ref{goftbl}, in relative terms. That is, in
Panel A, each fit measure for PFSBS is presented as a ratio to the
corresponding fit measure for the raw sieve method (SBS); hence, a value less
than one indicates that the pre-filtering yields a distribution that is a
better fit to the Monte Carlo-based distribution. The corresponding results
for FPFBS are presented in Panel B. Results recorded in all four figures and
both panels of the table are for $T=500.$

A visual inspection of the graphs in Figures \ref{bsmcfig1} and \ref{bsmcfig2}
suggests that when the long memory parameter is small $(d=0.2)$, and for both
values of $\phi$, the two bootstrap methods, SBS and PFSBS, provide reasonable
accuracy. There is no clear cut superiority of one method over the other,
with the raw sieve method being superior to the PFSBS method for $k=1$ and
$9$, and the opposite result obtaining for $k=3$ and $6.$ These visual results
on relative performance are confirmed (overall) by the numerical results in
Table \ref{goftbl}, Panel A, with virtually all ratios (associated with all
three measures of fit) being greater than one (indicating the superiority of
the raw method) for $k=1$ and $9,$ and less than one for $k=3$ and $6.$

In contrast, for $d=0.4$, the performances of the two methods are more
distinct, with all graphs reproduced in Figures \ref{bsmcfig3} and
\ref{bsmcfig4} - allied with the numerical results reported in Panel A, Table
\ref{goftbl} - confirming the marked superiority of the PFSBS method in this
part of the parameter space. Taken together these two sets of results suggest
that a conservative approach to estimating the sampling distribution of the
autocorrelation coefficients in empirical settings is to undertake the
pre-filtering; the increase in accuracy in the long memory region being worth
the slight reduction that \textit{may} occur (relative to the raw sieve) if
the true value of $d$ is small. %

\begin{table}[tbp] \centering
\caption{Goodness of fit measures for $\hat{\protect\rho}(k)$, $k=1,3,6,9$:\\
measures for PFSBS and FPFBS relative to those for SBS. \\
All results are for sample size $T=500$. } \label{goftbl}%

\begin{tabular}
[c]{ccccccccccc}
&  &  &  &  &  &  &  &  &  & \\
&  & \multicolumn{9}{c}{Panel A: PFSBS}\\
&  &  &  &  &  &  &  &  &  & \\
&  & \multicolumn{4}{c}{$d=0.2$} &  & \multicolumn{4}{c}{$d=0.4$}%
\\\cline{3-6}\cline{8-11}
&  &  &  &  &  &  &  &  &  & \\
&  & \multicolumn{4}{c}{Lag length $k$} &  & \multicolumn{4}{c}{Lag length $k
$}\\\cline{3-6}\cline{8-11}
&  & $1$ & $3$ & $6$ & $9$ &  & $1$ & $3$ & $6$ & $9$\\
& $\phi$ &  &  &  &  &  &  &  &  & \\\cline{2-2}%
RMSE & 0.3 & 4.566 & 0.384 & 0.616 & 1.276 &  & 0.235 & 0.173 & 0.358 &
0.374\\
& 0.6 & 1.467 & 0.520 & 0.682 & 1.343 &  & 0.322 & 0.233 & 0.392 & 0.387\\
KLD & 0.3 & 12.26 & 0.733 & 1.087 & 1.579 &  & 0.302 & 0.095 & 0.141 & 0.163\\
& 0.6 & 6.357 & 1.304 & 0.858 & 1.310 &  & 0.500 & 0.130 & 0.195 & 0.231\\
GINI & 0.3 & 9.006 & 0.339 & 0.605 & 1.128 &  & 0.264 & 0.128 & 0.234 &
0.333\\
& 0.6 & 2.679 & 0.418 & 0.611 & 1.211 &  & 0.370 & 0.220 & 0.290 & 0.349\\
&  &  &  &  &  &  &  &  &  & \\
&  & \multicolumn{9}{c}{Panel B: FPFBS}\\
&  &  &  &  &  &  &  &  &  & \\
&  & \multicolumn{4}{c}{$d=0.2$} &  & \multicolumn{4}{c}{$d=0.4$}%
\\\cline{3-6}\cline{8-11}
&  &  &  &  &  &  &  &  &  & \\
&  & \multicolumn{4}{c}{Lag length $k$} &  & \multicolumn{4}{c}{Lag length $k
$}\\\cline{3-6}\cline{8-11}
&  & $1$ & $3$ & $6$ & $9$ &  & $1$ & $3$ & $6$ & $9$\\
& $\phi$ &  &  &  &  &  &  &  &  & \\\cline{2-2}%
RMSE & 0.3 & 16.52 & 2.472 & 3.597 & 4.486 &  & 0.597 & 0.409 & 0.567 &
0.797\\
& 0.6 & 4.385 & 2.416 & 2.699 & 4.257 &  & 0.597 & 0.436 & 0.530 & 0.662\\
KLD & 0.3 & 102.99 & 10.32 & 18.79 & 18.18 &  & 0.971 & 0.380 & 0.426 &
0.507\\
& 0.6 & 34.030 & 13.110 & 9.644 & 11.15 &  & 1.129 & 0.447 & 0.425 & 0.491\\
GINI & 0.3 & 36.173 & 3.239 & 4.117 & 5.522 &  & 0.743 & 0.471 & 0.548 &
0.794\\
& 0.6 & 9.979 & 2.946 & 3.220 & 4.829 &  & 0.759 & 0.502 & 0.519 & 0.655\\
&  &  &  &  &  &  &  &  &  &
\end{tabular}%
\end{table}%

Turning, finally, to the FPFBS, from inspection of Figures \ref{bsmcfig1}%
-\ref{bsmcfig4} and the results recorded in Panel B of Table \ref{goftbl}, we
see that while it virtually always outperforms the raw SBS for the larger
value of $d$ ($d=0.4$), it does very poorly for $d=0.2$. In the latter case we
observe a \textquotedblleft divergence\textquotedblright\ (as measured by our
three goodness of fit\ measures) from the Monte Carlo distribution we are
attempting to replicate that is several times larger than that of SBS; more
than 100 times larger in one case. Most importantly, if we compare the FPFBS
\emph{directly} to the PFSBS (by making the appropriate simple calculations
using the numbers recorded in the two panels of Table \ref{goftbl}) we see
that the FPSBS \textit{never }outperforms the PFSBS, with the goodness of fit
measures for the former ranging from (approximately) twice to seventeen times
those of the latter. This poor performance (overall) of the FPFBS mimics that
documented for the sample mean.

\clearpage

\section{Summary and Conclusion}

This paper has derived new results regarding the convergence rates of
sieve-based bootstrap techniques, in the context of fractionally integrated
processes. Both the raw sieve technique, based on an autoregressive
approximation of the long memory process, and a pre-filtered version of the
sieve method, are investigated, for a broad class of statistics that includes
the sample mean and sample second-order moments. Pre-filtering via an
appropriate estimator is shown to yield a convergence rate that is equivalent
to that associated with intermediate and short memory processes, which is, in
turn, arbitrarily close to that associated with independent data.

Using numerical simulation, the distinct (and only rarely noted) problem of
underestimating the sampling variance of the sample mean in the long memory
case is shown to be avoided, in large measure, by use of a pre-filtering
method based, in turn, on a (bias-adjusted) semi-parametric estimator of the
long memory parameter. In particular, for moderate values of $d$ the
pre-filtered sieve produces very accurate estimates of the (known) exact
distribution of the sample mean, and achieves reasonable accuracy elsewhere in
the parameter space. Replacing the data-based pre-filter with a fixed value
may produce a slight improvement, but only when the latter is close to the
true parameter; otherwise the fixed pre-filtering performs very badly in terms
of reproducing the exact distribution.

The (data-based) pre-filtering technique is also shown to produce very
accurate estimates of the true sampling distribution of selected
autocorrelation coefficients (as measured by Monte Carlo simulation). Whilst
there is no clear cut superiority of the pre-filtered over the raw sieve
method when the fractional integration parameter is small, as the fractional
integration parameter increases the performance of the two methods becomes
more distinct, and the pre-filtering method performs notably better,
reflecting the properties established in the theoretical development. As is
the case with the sample mean, while fixed pre-filtering can outperform the
raw method when the assigned pre-filtering value is close to the true
parameter, it does very poorly otherwise, and in any case never outperforms
data-based pre-filtering in terms of reproducing the (Monte Carlo) sampling
distributions of the sample autocorrelations. Finally, for the narrow region
of the parameter space in which an Edgeworth approximation of the distribution
of the sample autocorrelations is valid, the sieve bootstrap reproduces this
analytical result with great accuracy.

With due acknowledgement made of the limited nature of the current
experimental exercise, we conclude that the overall increase in accuracy
obtained when using the data-based pre-filtered sieve bootstrap in parts of
the parameter space associated with moderate to strong long-range dependence
is worth the slight reduction that might occur (relative to the raw sieve)
otherwise; and that a reasonable approach to estimating unknown sampling
distributions in empirical settings is to employ the data-based pre-filtered
sieve as the default method.

\appendix%
\makeatletter\def\@seccntformat#1{\csname Pref@#1\endcsname: \ }
\def\Pref@section{Appendix}
\makeatother

\section{Edgeworth expansion for the sample autocorrelation function}

To support the reproducibility of the results reported in this paper, we
provide a brief outline of the details of the Edgeworth expansion used as a
comparator of our bootstrap-based methodology. All further details of this
expansion can be found in \cite{lieberman:rousseau:zucker:2001} (LRZ
hereafter).

Suppose we possess a statistic $s_{T}$ such that $\zeta_{T}=\sqrt{T}%
(s_{T}-E[s_{T}])\overset{\mathcal{D}}{\rightarrow}\mathbb{N}(0,\nu^{2})$. The
conventional (second-order) Edgeworth expansion for the CDF\ of $\zeta_{T}$ is
of the form
\begin{align}
F_{\zeta_{T}}(c)  &  =\Pr\left\{  \zeta_{T}<c=\nu u\right\} \label{Edge}\\
&  =G\left(  u\right)  -\left\{  \frac{\eta_{3}}{6\sqrt{T}}H_{2}\left(
u\right)  +\frac{1}{T}\left(  \frac{\eta_{4}}{24}H_{3}\left(  u\right)
+\frac{\eta_{3}^{2}}{72}H_{5}\left(  u\right)  \right)  \right\}  g\left(
u\right)  +O\left(  T^{-3/2}\right) \nonumber
\end{align}
where $\eta_{r}=\left.  \kappa_{r}\right/  \nu^{r}$ denotes the $r^{th}$
standardised cumulant of $\zeta_{T},$ and%
\begin{align*}
H_{2}\left(  u\right)   &  =u^{2}-1\\
H_{3}\left(  u\right)   &  =u^{3}-3u\\
H_{5}\left(  u\right)   &  =u^{5}-10u^{3}+15u
\end{align*}
are the required Hermite polynomials \citep{Hall:1992}. Accordingly, direct
application of \eqref{Edge} to the statistic
\begin{equation}
W_{T}(k)=\sqrt{T}\left(  \hat{\rho}(k)-\rho(k)\right)  \label{w}%
\end{equation}
requires a means of computing the required cumulants of $W$. In theory these
might be computed via \citet[Theorem 6]{magnus:1986}, or possibly
\cite{smith:1989}; in practice these expressions quickly become unmanageable
as the order of the required moments increases.

LRZ instead begin with
\begin{equation}
Z_{T}(k)=\sqrt{n}\left(  \hat{\gamma}(k)-\gamma(k)\right)  , \label{z}%
\end{equation}
where%
\begin{align*}
\widehat{\gamma}(k)  &  =\frac{1}{T}\sum_{t=1}^{T-k}x(t)x(T+k)=T^{-1}%
x^{\prime}A_{T,k}x\,,\\
\left[  A_{T,k}\right]  _{i,j}  &  =\xi_{k}\left(  i-j\right)  =\left\{
\begin{array}
[c]{cl}%
\frac12
& \text{for}\quad\left\vert i-j\right\vert =k\\
0 & \text{otherwise}%
\end{array}
\right.  ,
\end{align*}
and $A_{T,0}=I_{T}$. Assuming that $x=(x(1),\ldots,x(T))^{\prime}$ is
distributed $\mathbb{N}\left(  0,\Sigma\right)  $\footnote{LRZ explicitly
impose $E(X)=0$; or, equivalently, assume that $X=Y-\mu$ where $\mu=E(Y)$ is
known.}, LRZ then proceed to produce an expansion for $W_{T}(k)$ indirectly,
via $Z_{T}(k)$, as follows.

For brevity write $Z_{T}(k)$ and $W_{T}(k)$ as $Z_{k}$ and $W_{k}$
respectively, and use $\rho(k)=\gamma(k)/\gamma(0)$ and $\hat{\gamma
}(k)=\gamma(k)+T^{-1/2}Z_{k}$ to now rewrite $W_{k}$ as%
\begin{align*}
W_{k}  &  =\sqrt{T}\left(  \frac{\hat{\gamma}(k)}{\hat{\gamma}(0)}%
-\frac{\gamma(k)}{\gamma(0)}\right) \\
&  =\sqrt{T}\left(  \frac{\hat{\gamma}(k)\gamma(0)-\gamma(k)\hat{\gamma}%
(0)}{\hat{\gamma}(0)\gamma(0)}\right) \\
&  =\frac{\gamma(0)Z_{k}-\gamma(k)Z_{0}}{\gamma(0)\hat{\gamma}(0)}\\
&  =\frac{Z_{k}-\rho(k)Z_{0}}{\hat{\gamma}(0)}.
\end{align*}
Then for a single $W_{k}:$%
\begin{align*}
F_{W_{k}}(c)  &  =\Pr\left(  W_{k}<c\right) \\
&  =\Pr\left(  Z_{k}-\rho(k)Z_{0}<c\hat{\gamma}(0)\right)  \quad\text{since
}\hat{\gamma}(0)>0\\
&  =\Pr\left(  Z_{k}-\rho(k)Z_{0}<c\gamma(0)+cT^{-1/2}Z_{0}\right) \\
&  =\Pr\left(  Z_{k}-\left(  \rho(k)+cT^{-1/2}\right)  Z_{0}<c\gamma
(0)\right)  ,
\end{align*}
where, from (\ref{z}),%
\begin{align*}
Z_{k}-\left(  \rho(k)+cT^{-1/2}\right)  Z_{0}  &  =\sqrt{T}\left(  \hat
{\gamma}(k)-\left(  \rho(k)+cT^{-1/2}\right)  \hat{\gamma}(0)\right) \\
&  -\left[  \sqrt{T}\left(  \gamma(k)-\left(  \rho(k)+cT^{-1/2}\right)
\gamma(0)\right)  \right] \\
&  =\sqrt{T}\left(  \hat{\gamma}(k)-\left(  \rho(k)+cT^{-1/2}\right)
\hat{\gamma}(0)\right)  +c\gamma(0)\\
&  =\sqrt{T}\left(  n^{-1}x^{\prime}A_{T,k}x-\left(  \rho(k)+cT^{-1/2}\right)
T^{-1}x^{\prime}x\right)  +c\gamma(0)\\
&  =T^{-1/2}x^{\prime}B_{T,k}x+c\gamma(0)
\end{align*}
and $B_{T,k}=A_{T,k}-\left(  \rho(k)+cT^{-1/2}\right)  I_{T}.$ So, defining
$x^{\prime}B_{T,k}x=Q_{T,k}^{\dag},$ we have
\[
F_{W_{k}}(c)=\Pr\left(  W_{k}<c\right)  \equiv\Pr\left\{  Q_{T,k}^{\dag
}<0\right\}  .
\]

Standard results on quadratic forms in normal variates when $x\sim
\mathbb{N}(0,\Sigma)$ gives $\psi_{T}(\tau)=E[i\tau Q_{T,k}^{\dag}%
]=\prod_{t=1}^{T}(1-2i\tau\lambda_{t})^{-\half}$ on application of Aitken's
integral, where $\lambda_{t}$, $t=1,\ldots,T$ are the eigenvalues of
$B_{T,k}\mathbf{\Sigma}$. The characteristic function is integrable for all
$T>2$ and the cumulant generating function $-\half\sum_{t=1}^{T}\log
(1-2i\tau\lambda_{t})$ yields the $r^{th}$ cumulant of $Q_{T,k}^{\dag}$ as
\[
\kappa_{r}^{\ast}=2^{r-1}\left(  r-1\right)  !\mathrm{tr}\left[  \left(
B_{T,k}\mathbf{\Sigma}\right)  ^{r}\right]  \,.
\]
Evaluating the mean and variance now makes $Q_{T,k}^{\dag}$ (or more correctly
its $z$-score) a convenient candidate for an Edgeworth expansion.

From the preceding,
\[
F_{W_{k}}(c)\equiv\Pr\left\{  Q_{T,k}^{\dag}<0\right\}  =\Pr\left\{
\frac{Q_{T,k}^{\dag}-\mu^{\dag}}{\sigma^{\dag}}<u=-\frac{\mu^{\dag}}%
{\sigma^{\dag}}\right\}  ,
\]
where $\mu^{\dag}=\kappa_{1}^{\dag}$ and $\sigma^{\dag}=\sqrt{\kappa_{2}%
^{\dag}}$. Hence the second-order Edgeworth expansion (if it exists) for the
CDF of $Q_{T,k}^{\dag}$ (and hence $W_{k})$ will be of the form\footnote{It
should be noted that LRZ give the expansion for $1-F_{W_{k}}(c),$ rather than
$F_{W_{k}}(c)$ itself.}%
\[
\widetilde{F}_{W_{k}}(c)=G\left(  u\right)  -\left\{  \frac{\eta_{3}^{\dag}%
}{6}H_{2}\left(  u\right)  +\frac{\eta_{4}^{\dag}}{24}H_{3}\left(  u\right)
+\frac{(\eta_{3}^{\dag})^{2}}{72}H_{5}\left(  u\right)  \right\}  g\left(
u\right)  ,
\]
with error $O\left(  T^{-3/2}\right)  ,$ where $\eta_{r}^{\dag}=\kappa
_{r}^{\dag}/(\sqrt{\kappa_{2}^{\dag}})^{r}$, $r=1,2,3,4$, and $u=-\eta
_{1}^{\dag}$. Note that the descending powers of $\sqrt{T}$ that would
ordinarily appear in the expansion (\textit{cf}. \eqref{Edge}) are here
subsumed into the standardised cumulants; that is, we are implicitly assuming
that $\eta_{r}^{\dag}=O\left(  T^{1-r/2}\right)  $ or, equivalently, that
$\kappa_{r}^{\dag}$ is $O\left(  T\right)  $, at least up to $r=4$.

That the cumulants of $Q_{T,k}^{\dag}$ are of the appropriate order, at least
for restricted values of the fractional parameter $d,$ follows from LRZ
Theorem 1. In particular, the cumulants of $Q_{T,k}^{\dag}$ of order no
greater than $r$ will be $O\left(  T\right)  $ only if $r(2d)<1$, implying
that $\kappa_{r}^{\dag}$, $r=1,2,3,4$, are $O\left(  T\right)  $ if $d<0.125$
but not otherwise. However, if $r$ now denotes the order of the highest
cumulant in the expansion, then LRZ also show that we require $(r+1)(2d)<1$
and $r$ even to attain an expansion error of $o\left(  T^{1-r/2}\right)  $;
while if $r$ is odd the error is of the same order as the last term, namely
$O\left(  T^{1-r/2}\right)  $. Hence the second-order ($r=4$) expansion is
valid only for $d<0.1$, and there is no valid expansion (in the sense that the
error is of smaller order than the last included term) for $d\geq0.1$.

\bibliographystyle{ims}
\bibliography{tsa}

\begin{thebibliography}{41}
\expandafter\ifx\csname natexlab\endcsname\relax\def\natexlab#1{#1}\fi

\bibitem[{Adenstedt(1974)}]{adenstedt:1974}
\textsc{Adenstedt, R.~K.} (1974).
\newblock On large--sample estimation for the mean of a stationary sequence.
\newblock \textit{The Annals of Statistics} \textbf{2} 1095--1107.

\bibitem[{Andrews et~al.(2006)Andrews, Lieberman and
  Marmer}]{andrews:lieberman:marmer:2006}
\textsc{Andrews, D.~W.}, \textsc{Lieberman, O.} and \textsc{Marmer, V.} (2006).
\newblock Higher-order improvements of the parametric bootstrap for long-memory
  {G}aussian processes.
\newblock \textit{Journal of Econometrics} \textbf{133} 673--702.

\bibitem[{Andrews and Lieberman(2005)}]{andrews:lieberman:2005}
\textsc{Andrews, D. W.~K.} and \textsc{Lieberman, O.} (2005).
\newblock Valid edgeworth expansions for the {W}hittle maximum likelihood
  estimator for stationary long-memory gaussian time series.
\newblock \textit{Econometric Theory} \textbf{21} 710--734.

\bibitem[{Andrews and Sun(2004)}]{andrew:sun:2004}
\textsc{Andrews, D. W.~K.} and \textsc{Sun, Y.} (2004).
\newblock Adaptive local polynomial {W}hittle estimation of long-range
  dependence.
\newblock \textit{Econometrica} \textbf{72} 569--614.

\bibitem[{Apostol(1960)}]{apostol:1960}
\textsc{Apostol, T.~M.} (1960).
\newblock \textit{Mathematical Analysis}.
\newblock Addison-Wesley, Reading.

\bibitem[{Beran(1994)}]{beran:1994}
\textsc{Beran, J.} (1994).
\newblock \textit{Statistics for long-memory processes}, vol.~61 of
  \textit{Monographs on Statistics and Applied Probability}.
\newblock Chapman and Hall, New York.

\bibitem[{Beran(1995)}]{beran:1995}
\textsc{Beran, J.} (1995).
\newblock Maximum likelihood estimation of the differencing parameter for
  invertible short and long memory autoregressive integrated moving average
  models.
\newblock \textit{Journal of the Royal Statistical Society} \textbf{{\bf B} 57}
  654--672.

\bibitem[{Bickel and Freedman(1981)}]{bickel:freedman:1981}
\textsc{Bickel, P.~J.} and \textsc{Freedman, D.~A.} (1981).
\newblock Some asymptotic theory for the bootstrap.
\newblock \textit{Annals of Statistics} \textbf{9} 1196--1217.

\bibitem[{Brockwell and Davis(1991)}]{brockwell:davis:1991}
\textsc{Brockwell, P.~J.} and \textsc{Davis, R.~A.} (1991).
\newblock \textit{Time Series: Theory and Methods}.
\newblock Springer Series in Statistics. Springer-Verlag, New York, 2nd ed.

\bibitem[{B\"{u}hlmann(1997)}]{buhlmann:1997}
\textsc{B\"{u}hlmann, P.} (1997).
\newblock Sieve bootstrap for time series.
\newblock \textit{Bernoulli} \textbf{3} 123--148.

\bibitem[{Choi and Hall(2000)}]{choi:hall:2000}
\textsc{Choi, E.} and \textsc{Hall, P.~G.} (2000).
\newblock Bootstrap confidence regions from autoregressions of arbitrary order.
\newblock \textit{Journal of the Royal Statistical Society} \textbf{{\bf B} 62}
  461--477.

\bibitem[{Dahlhaus(1989)}]{dahlhaus:1989}
\textsc{Dahlhaus, R.} (1989).
\newblock Efficient parameter estimation for self-similar processes.
\newblock \textit{Annals of Statistics} \textbf{17} 1749--1766.

\bibitem[{Doornik and Ooms(2001)}]{doornik:ooms:2003}
\textsc{Doornik, J.~A.} and \textsc{Ooms, M.} (2001).
\newblock Computational aspects of maximum likelihood estimation of
  autoregressive fractionally integrated moving average models.
\newblock \textit{Computational Statistics \& Data Analysis} \textbf{42}
  333--348.
\newblock Also a 2001 Nuffield discussion paper.

\bibitem[{Durbin(1980)}]{durbin:1980}
\textsc{Durbin, J.} (1980).
\newblock Approximations for densities of sufficient estimators.
\newblock \textit{Biometrika} \textbf{67} 311--333.

\bibitem[{Fox and Taqqu(1986)}]{fox:taqqu:1986}
\textsc{Fox, R.} and \textsc{Taqqu, M.~S.} (1986).
\newblock Large sample properties of parameter estimates for strongly dependent
  stationary gaussian time series.
\newblock \textit{Annals of Statistics} \textbf{14} 517--532.

\bibitem[{Geweke and Porter-Hudak(1983)}]{geweke:porter:1983}
\textsc{Geweke, J.} and \textsc{Porter-Hudak, S.} (1983).
\newblock The estimation and application of long-memory time series models.
\newblock \textit{Journal of Time Series Analysis} \textbf{4} 221--238.

\bibitem[{Giraitis and Robinson(2003)}]{giraitis:robinson:2003}
\textsc{Giraitis, L.} and \textsc{Robinson, P.~M.} (2003).
\newblock Edgeworth expansions for semiparametric {W}hittle estimation of long
  memory.
\newblock \textit{Annals of Statistics} \textbf{31} 1325--1375.

\bibitem[{Granger and Joyeux(1980)}]{granger:joyeux:1980}
\textsc{Granger, C. W.~J.} and \textsc{Joyeux, R.} (1980).
\newblock An introduction to long-memory time series models and fractional
  differencing.
\newblock \textit{Journal of Time Series Analysis} \textbf{1} 15--29.

\bibitem[{Hall(1992)}]{Hall:1992}
\textsc{Hall, P.} (1992).
\newblock \textit{The bootstrap and {E}dgeworth expansion}.
\newblock Springer series in Statistics. Springer-Verlag, New York.

\bibitem[{Hesterberg(1997)}]{hesterberg:1997}
\textsc{Hesterberg, T.} (1997).
\newblock Matched-block bootstrap for long memory processes.
\newblock Research Report~66, MathSoft, Inc, Seattle, WA.

\bibitem[{Hosking(1980)}]{hosking:1981}
\textsc{Hosking, J. R.~M.} (1980).
\newblock Fractional differencing.
\newblock \textit{Biometrika} \textbf{68} 165--176.

\bibitem[{Hosking(1996)}]{hosking:1996}
\textsc{Hosking, J. R.~M.} (1996).
\newblock Asymptotic distributions of the sample mean, autocovariances, and
  autocorrelations of long memory time series.
\newblock \textit{Journal of Econometrics} \textbf{73} 261--284.

\bibitem[{Inoue and Kasahara(2006)}]{inoue:kasahara:2006}
\textsc{Inoue, A.} and \textsc{Kasahara, Y.} (2006).
\newblock Explicit representation of finite predictor coefficients and its
  applications.
\newblock \textit{Annals of Statistics} \textbf{34} 973--993.

\bibitem[{Kreiss et~al.(2011)Kreiss, Paparoditis and
  Politis}]{kreiss:paparoditis:politis:2011}
\textsc{Kreiss, J.~P.}, \textsc{Paparoditis, E.} and \textsc{Politis, D.~N.}
  (2011).
\newblock On the range of validity of the autoregressive sieve bootstrap.
\newblock \textit{Annals of Statistics} \textbf{39} 2103--2130.

\bibitem[{K\"{u}nsch(1989)}]{kunsch:1989}
\textsc{K\"{u}nsch, H.~R.} (1989).
\newblock The jacknife and the bootstrap for general stationary observations.
\newblock \textit{Annals of Statistics} \textbf{17} 1217--1241.

\bibitem[{Lieberman and Phillips(2004)}]{lieberman:phillips:2004}
\textsc{Lieberman, O.} and \textsc{Phillips, P. C.~B.} (2004).
\newblock Expansions for the distribution of the maximum likelihood estimator
  of the fractional difference parameter.
\newblock \textit{Econometric Theory} \textbf{20} 464--484.

\bibitem[{Lieberman et~al.(2001)Lieberman, Rousseau and
  Zucker}]{lieberman:rousseau:zucker:2001}
\textsc{Lieberman, O.}, \textsc{Rousseau, J.} and \textsc{Zucker, D.~M.}
  (2001).
\newblock Valid {E}dgeworth expansion for the sample autocorrelation function
  under long range dependence.
\newblock \textit{Econometric Theory} \textbf{17} 257--275.

\bibitem[{Lieberman et~al.(2003)Lieberman, Rousseau and
  Zucker}]{lieberman:rousseau:zucker:2003}
\textsc{Lieberman, O.}, \textsc{Rousseau, J.} and \textsc{Zucker, D.~M.}
  (2003).
\newblock Valid asymptotic expansions for the maximum likelihood estimator of
  the parameter of a stationary, gaussian, strongly dependent process.
\newblock \textit{The Annals of Statistics} \textbf{31} 586--612.

\bibitem[{Magnus(1986)}]{magnus:1986}
\textsc{Magnus, J.~R.} (1986).
\newblock The exact moments of a ratio of quadratic forms in normal variables.
\newblock \textit{Annals of Economics and Statistics / Annales d'Économie et de
  Statistique}  pp. 95--109.

\bibitem[{Nielsen and Frederiksen(2005)}]{nielsen:frederiksen:2005}
\textsc{Nielsen, M.~.} and \textsc{Frederiksen, P.~H.} (2005).
\newblock Finite sample comparison of parametric, semiparametric, and wavelet
  estimators of fractional integration.
\newblock \textit{Econometric Reviews} \textbf{24} 405--443.

\bibitem[{Politis(2003)}]{politis:2003}
\textsc{Politis, D.~N.} (2003).
\newblock The impact of bootstrap methods on time series analysis.
\newblock \textit{Statistical Science}  219--230.

\bibitem[{Poskitt(1994)}]{poskitt:1994}
\textsc{Poskitt, D.~S.} (1994).
\newblock A note on autoregressive modelling.
\newblock \textit{Econometric Theory} \textbf{10} 884--899.

\bibitem[{Poskitt(2007)}]{poskitt:2007}
\textsc{Poskitt, D.~S.} (2007).
\newblock Autoregressive approximation in nonstandard situations: The
  fractionally integrated and non-invertible cases.
\newblock \textit{Annals of Institute of Statistical Mathematics} \textbf{59}
  697--725.

\bibitem[{Poskitt(2008)}]{poskitt:2008}
\textsc{Poskitt, D.~S.} (2008).
\newblock Properties of the sieve bootstrap for fractionally integrated and
  non-invertible processes.
\newblock \textit{Journal of Time Series Analysis} \textbf{29} 224--250.

\bibitem[{Poskitt et~al.(2012)Poskitt, Martin and
  Grose}]{poskitt:martin:grose:2012}
\textsc{Poskitt, D.~S.}, \textsc{Martin, G.~M.} and \textsc{Grose, S.~G.}
  (2012).
\newblock Bias reduction of long memory parameter estimators via the
  pre-filtered sieve bootstrap.
\newblock Econometrics \& Business Statistics Working Paper WP 08/12, Monash
  University.

\bibitem[{Robinson(1995{\natexlab{a}})}]{robinson:1995b}
\textsc{Robinson, P.~M.} (1995{\natexlab{a}}).
\newblock Gaussian semiparametric estimation of long range dependence.
\newblock \textit{Annals of Statistics} \textbf{23} 1630--1661.

\bibitem[{Robinson(1995{\natexlab{b}})}]{robinson:1995a}
\textsc{Robinson, P.~M.} (1995{\natexlab{b}}).
\newblock Log periodogram regression of time series with long memory.
\newblock \textit{Annals of Statistics} \textbf{23} 1048--1072.

\bibitem[{Shibata(1980)}]{shibata:1980}
\textsc{Shibata, R.} (1980).
\newblock Asymptotically efficient selection of the order of the model for
  estimating parameters of a linear process.
\newblock \textit{Annals of Statistics} \textbf{8} 147--164.

\bibitem[{Smith(1989)}]{smith:1989}
\textsc{Smith, M.~D.} (1989).
\newblock On the expectation of a ratio of quadratic forms in normal variables.
\newblock \textit{Journal of Multivariate Analysis} \textbf{31} 244--257.

\bibitem[{Sowell(1992)}]{sowell:1992}
\textsc{Sowell, F.} (1992).
\newblock Maximum likelihood estmation of stationary univariate fractionally
  integrated time series models.
\newblock \textit{Journal of Econometrics} \textbf{53} 165--188.

\bibitem[{Taniguchi(1984)}]{taniguchi:1984}
\textsc{Taniguchi, M.} (1984).
\newblock Validity of {E}dgeworth expansions for statistics of time series.
\newblock \textit{Journal of Time Series Analysis} \textbf{5} 37--51.

\end{thebibliography}

\end{document}